\documentclass[12pt]{article}
\usepackage{openwork}
\usepackage{xcolor}
\usepackage{tikz}
\usetikzlibrary{arrows.meta, positioning, calc}
\usepackage{setspace}
\RequirePackage{float}
\usepackage{hyperref}
\usepackage{pifont}
\newcommand{\cmark}{\ding{51}} 

\title{The Laziness of the Crowd: Effort Aversion Among Raters Risks Undermining the Efficacy of X's Community Notes Program}
\author[1,2]{Morgan Wack\thanks{Correspondence: \texttt{m.wack@ikmz.uzh.ch}}}
\author[2,3]{Patrick Warren}
\author[2,3]{Mustafa Alam}
\affil[1]{\textit{Department of Communication and Media Research (IKMZ), University of Zurich, Switzerland}}
\affil[2]{\textit{Media Forensics Hub, Clemson University, USA}}
\affil[3]{\textit{John E. Walker Department of Economics, Clemson University, USA}}
\date{\today}

\begin{document}

\maketitle

\begin{abstract}
\textit{Crowdsourced moderation systems like Twitter/X's Community Notes program have been proposed as scalable alternatives to professional fact-checkers for combating online misinformation. While prior research has examined the effectiveness of such systems in reducing engagement with false content and their vulnerability to partisan bias, we identify a previously untested mechanism linking fact-check difficulty to systematic non-participation by crowdsourced raters. We hypothesize that claims requiring less cognitive effort to evaluate, specifically, those that are obviously false and easy to refute, are more likely to receive public notes than claims that are more plausible and require greater effort to debunk. Using eighteen months of vaccine-related Community Notes data (2,250 posts) and ratings from 382 survey participants, we show that claims perceived as more difficult to fact-check are significantly less likely to receive notes that achieve ``helpful''/public status. Following the conduct of additional analyses and a fact-checking process utilizing an LLM pipeline to help rule out alternative explanations, we interpret this pattern as consistent with an unwillingness among raters to invest the mental effort required to evaluate and rate notes for more plausible misinformation. These findings suggest that crowdsourced moderation may systematically fail to address the forms of plausible misinformation which are most likely to deceive. We discuss implications for platform design and propose mechanisms to mitigate this difficulty penalty in crowdsourced content moderation systems.}
\end{abstract}

\clearpage
\onehalfspacing
\section{Introduction}

The proliferation of false information on social media platforms poses significant threats to public health, democratic processes, and social cohesion \parencite{ecker2024misinformation, pennycook2020fighting}, with far-reaching consequences for the epistemic quality of public discourse \parencite{lazer2018science, lewandowsky2012misinformation}. As online misinformation often spreads faster and reaches more people than accurate information online \parencite{vosoughi2018spread}, researchers and platform operators have devoted considerable attention to developing scalable interventions. Professional fact-checking has long served as the primary mechanism for identifying and labeling false content, demonstrating meaningful reductions in belief and sharing of misinformation when warnings are attached to posts \parencite{martel2024crowds}. However, the speed and volume of content creation on social media exceeds the capacity of professional fact-checkers, leaving many false claims unchecked or checked only after their initial spread \parencite{wack2024political}. Furthermore, growing public skepticism toward fact-checking organizations, particularly among conservative-leaning audiences who perceive liberal bias, has diminished the effectiveness of these expert-driven interventions \parencite{walker2019republicans}.

In response to these limitations, platforms have increasingly turned to crowdsourced moderation systems as scalable alternatives. X/Twitter's Community Notes program, originally launched as Birdwatch in 2021, represents the most prominent implementation of this approach. The system enables ordinary users to propose contextual notes for potentially misleading posts and rate the helpfulness of proposed notes submitted by others. An algorithm designed to identify notes that achieve consensus across ideologically diverse raters determines which notes are displayed publicly \parencite{wojcik2022birdwatch}. The theoretical foundation for this approach rests on the ``wisdom of the crowd'' principle which suggests that aggregated judgments from groups of non-experts can produce assessments comparable in accuracy to professional fact-checkers \parencite{galton1907vox}. 

Empirical evaluations have generated encouraging results. Community Notes reduce subsequent reposts by approximately 46\% and likes by 44\% when applied \parencite{slaughter2025community}, and exposure to notes reduces users' agreement with the substance of misleading posts \parencite{wojcik2022birdwatch}. Despite this evidence, the system faces serious limitations. Over a period of four years, only 13.55\% of posts with at least one proposed Note ever became public noted \parencite{mohammadi2025birdwatch}, and simulation studies suggest the bridging algorithm suppresses over 40\% of genuinely helpful notes while remaining vulnerable to coordinated manipulation by as few as 5--20\% of raters \parencite{truong2025community}.

The present study introduces a new dimension to our understanding of crowdsourced moderation limitations. While prior work has examined partisan bias, manipulation vulnerability, and timeliness constraints, we propose that the cognitive demands of evaluating misinformation create systematic biases in which claims receive public notes. Specifically, we hypothesize that community notes are more likely to be applied to claims that are obviously false and perceived as easier to investigate, rather than to claims that are less obviously false and require more cognitive effort to fact-check. This pattern would present a troubling failure as it would mean that one of the most dangerous forms of misinformation, plausible claims that can readily deceive audiences, may be precisely the content least likely to receive corrective annotations at scale.

Our theoretical framework draws on the ``lazy, not biased'' account of misinformation susceptibility \parencite{pennycook2019lazy}, which demonstrates that people fall for false claims not primarily because of partisan motivation but because they fail to engage in effortful evaluation. We extend this insight from news consumers to the fact-checkers themselves: just as individuals accept plausible misinformation due to insufficient cognitive engagement, volunteer raters in crowdsourced systems may be less willing to invest the effort required to evaluate notes addressing nuanced false claims. The result would be a systematic under-representation of ratings for precisely the misinformation that requires correction the most.

Using eighteen months of vaccine-related Community Notes data comprising approximately 2,250 posts and crowdsourced judgments from 382 participants about their plausibility and fact-checking difficulty, we test whether the perceived fact-check difficulty of false claims predicts the likelihood of proposed notes achieving helpful status. We show that a critical form of misinformation, plausible claims that can readily deceive audiences, represents precisely the content least likely to receive corrective annotations. This concern is compounded by the work of \textcite{jalbert2026misaligned}, who find that removing plausible false content is uniquely effective at reducing conspiracy belief, yet lay participants systematically prioritize moderating \emph{implausible} content. Where similar intuitions guide Community Notes raters, crowdsourced systems may fail to correct precisely the misinformation whose removal would be most beneficial.

Finally, the urgency of understanding crowdsourced fact-checking has intensified as the approach spreads beyond X/Twitter. In the past two years, YouTube, Meta, and TikTok have all announced or launched Community Notes-style systems \parencite{lloyd2025beyond}. As a result, if left unaddressed, the difficulty penalty we document may soon affect content moderation across platforms serving billions of users.

\section{Literature Review}

As digital misinformation became a prominent public concern, professional fact-checking emerged as the dominant approach to combating misinformation, with warning labels attached to content verified as false by expert organizations. This emphasis was not unwarranted, with meta-analytic evidence indicating that when applied, fact-check labels effectively reduce belief in and willingness to share misinformation \parencite{walter2020factchecking}. The mechanism appears to operate through both direct correction of specific false beliefs and broader accuracy priming that makes individuals more discerning in subsequent information processing \parencite{pennycook2021shifting}. However, professional fact-checking efforts confront severe scalability constraints. The volume of potentially misleading content posted often exceeds fact-checking capacity, with most claims never receiving professional verification \parencite{stencel2021factchecking}. Moreover, the time required for careful fact-checking means that verdicts typically arrive after misinformation has already achieved wide circulation \parencite{wack2024political}. These structural limitations have motivated interest in alternative approaches that can operate at the scale of modern social media platforms.

Crowdsourced fact-checking offers a potential solution to the scalability problem by distributing evaluation across large numbers of platform users. The theoretical basis for this approach extends from early demonstrations of collective intelligence \parencite{galton1907vox, surowiecki2004wisdom} to more recent evidence that aggregated lay judgments can match expert performance across diverse domains \parencite{becker2017network}. In the specific context of misinformation, \textcite{allen2021scaling} found that groups of as little as twelve politically balanced laypeople could achieve accuracy levels comparable to professional fact-checkers when evaluating news headlines. Crucially, this accuracy held even for contentious political content, suggesting that crowd wisdom can overcome individual partisan biases when appropriately aggregated. Subsequent work confirmed these findings across different platforms, stimulus sets, and international contexts \parencite{arechar2023understanding, labarbera2024crowd, martel2024crowds}.

The implementation of crowdsourced fact-checking in practice, however, has revealed complications not fully captured in controlled experiments. \textcite{allen2022birds} examined early data from X/Twitter's Birdwatch program and found that contributors predominantly chose to fact-check posts from ideological opponents. While the content of notes produced remained accurate despite this selection bias, the finding raises questions about coverage. \textcite{saeed2022crowdsourced} compared crowdsourced notes to expert fact-checks and found that while crowds could match expert accuracy in some settings, they drew on inconsistent sources and failed to produce actionable results in others. These studies suggest that the controlled conditions producing favorable results in experiments may not fully translate to the messy reality of voluntary participation.

Community Notes specifically has attracted substantial research attention since its public launch. The system's bridging algorithm attempts to address partisan concerns by requiring consensus across ideologically diverse raters before displaying notes \parencite{wojcik2022birdwatch}. Notes achieving ``helpful'' status (and becoming publicly posted) must receive favorable ratings from contributors across the political spectrum, theoretically ensuring that only broadly accepted corrections gain visibility. \textcite{wojcik2022birdwatch} reported that exposure to bridging-selected notes reduced individual decisions to like and repost misinformation by 25-34\% relative to control conditions. Subsequent experimental work clarified the mechanism: textual community notes explaining why a post is misleading are perceived as significantly more trustworthy than simple misinformation flags, with the increased trustworthiness stemming primarily from the provided fact-checking explanations rather than from generally higher trust toward community fact-checkers \parencite{drolsbach2024trust}. This finding underscores what is lost when plausible misinformation fails to receive notes. Users are deprived not merely of a warning label but of the contextual explanation that makes corrections persuasive. 

Independent analyses have largely corroborated these effectiveness estimates while identifying important nuances. \textcite{slaughter2025community} employed synthetic control methods to estimate that note attachment reduces reposts by 46\% and likes by 44\% in the 48 hours following attachment. Notably, the magnitude of effects varied substantially with timing. Notes attached within 12 hours of post creation produced repost reductions of approximately 50\%, while notes attached after 47 hours showed minimal effects. This pattern underscores the critical importance of timeliness for intervention effectiveness.

The timeliness problem emerges as perhaps the most fundamental challenge facing Community Notes. \textcite{mohammadi2025birdwatch} report that helpful notes take an average of 26 hours to appear, well beyond the typical half-life of post engagement \parencite{pfeffer2023sample}. Analysis from the \textcite{ccdh2024communitynotes} found that 74\% of accurate notes related to the 2024 U.S. presidential election were never shown to users, allowing misleading posts without notes to accumulate 13 times more views than those eventually annotated. The bridging algorithm's requirement for cross-ideological consensus appears to contribute to these delays. Specifically, more polarizing content, which may be most in need of correction, faces higher barriers to achieving helpful status \parencite{mohammadi2025birdwatch}. Recent work by \textcite{wu2025beyond} documents a median delay of nearly 18 hours before health-related notes receive any helpfulness status, with approximately 88\% of notes never obtaining enough ratings to reach any determination.

Beyond timeliness, simulation research has exposed deeper vulnerabilities in the Community Notes algorithm. \textcite{truong2025community} conducted systematic evaluations using simulated rating data calibrated to match empirical distributions. Their analysis quantifies four types of potential errors, including suppression of helpful notes, pollution of published notes with unhelpful content, infiltration of unhelpful notes into the published set, and waste of helpful notes that go unpublished. Under realistic assumptions about rater distributions, suppression rates consistently exceeded 40\%. As a result, more than four in ten genuinely helpful notes failed to achieve publication. The algorithm also proved highly sensitive to rater polarization and in-group bias, with suppression and pollution rates reaching 80\% under conditions modeling realistic social divisions. Most concerning, their manipulation analysis revealed that coordinated bad-faith raters comprising just 5-20\% of the population could completely suppress helpful notes targeting content with particular characteristics.

Understanding why crowdsourced systems may under-perform requires engagement with the psychological literature on cognitive processing of information. Dual-process theories distinguish between fast, intuitive System 1 processing and slow, analytical System 2 processing \parencite{kahneman2011thinking, evans2013dual}. Extending these frameworks to misinformation, two competing accounts have emerged. The motivated reasoning account holds that individuals engage effortful cognition selectively to defend beliefs aligned with their identities and worldviews, rationalizing misinformation that supports their side while critically scrutinizing opposing content \parencite{kahan2017misconceptions}. The classical reasoning account, by contrast, predicts that analytical thinking supports accurate belief formation regardless of content alignment with prior attitudes \parencite{pennycook2019lazy}.

The influential study ``lazy, not biased'' from \textcite{pennycook2019lazy} provided strong evidence for the classical reasoning account in the context of fake news. Across multiple studies with over 3,400 participants, they found that performance on the Cognitive Reflection Test (CRT), a measure of propensity to engage analytical rather than intuitive thinking, negatively predicted perceived accuracy of fake news and positively predicted ability to discriminate fake from real news. Crucially, these relationships held even for headlines aligned with participants' political ideology. The results suggest that people often fall for misinformation not primarily because of partisan motivation but because they fail to engage in sufficient cognitive effort. Subsequent research extended these findings to misinformation sharing, showing that simply prompting users to consider accuracy before sharing reduces the spread of false content \parencite{pennycook2021shifting}. Additional replications have largely supported the laziness account across different national contexts, though some find that partisanship moderates the relationship between analytic thinking and misinformation discernment \parencite{farago2023hungarian}.  

The relationship between claim characteristics and fact-checking success has received limited direct attention, but related research offers suggestive evidence. \textcite{pennycook2020fighting} found that the correlation between cognitive reflection and disbelief in fake news was stronger for content that was more obviously implausible. This suggests that analytical thinking primarily helps with content where the falsity becomes apparent upon reflection, precisely the category we hypothesize receives disproportionate fact-checking attention. Studies of professional fact-checker selection also reveal systematic biases in which claims receive attention, with factors including political salience, claim specificity, and checkability influencing verification decisions \parencite{graves2018understanding}. If similar selection dynamics operate in crowdsourced systems, claims that are easier to evaluate may attract more rater engagement regardless of their relative importance. Our findings extend the 'cognitive miser' framework, which refers to the tendency of individuals to default to low-effort heuristics even when higher-order processing is required \parencite{fiske1991social, stanovich2018miserliness}. While previous work identifies this as a source of individual susceptibility to fake news, we demonstrate that it manifests as a collective blind spot, where the 'wisdom of crowds' is constrained by a 'laziness of crowds' with regards to high-effort verification.

The impact of fact-check difficulty we propose represents a novel synthesis of this prior work. We suggest that cognitive laziness, the same tendency that leads individuals to accept plausible misinformation rather than engaging in effortful evaluation, also affects which claims receive crowdsourced fact-checks. Notes addressing obviously false, easily refuted claims will receive faster and more favorable ratings because evaluation requires minimal cognitive investment. Notes addressing more plausible claims that require careful assessment of evidence will face higher barriers to achieving helpful status, as fewer raters invest the effort required to evaluate them properly. As a consequence, this could result in systematic coverage gaps in coverage for the types of plausible misinformation which most effectively reduces conspiratorial beliefs when removed \parencite{jalbert2026misaligned}.  

Several features of the Community Notes system may exacerbate this mechanism. First, the voluntary nature of rater participation means that cognitive effort represents a real cost that participants can avoid by simply not rating particular notes. Unlike experimental contexts where participants must evaluate all stimuli, crowdsourced raters can selectively engage with content that feels tractable. Beyond the cognitive costs of evaluating complex claims, raters may also undervalue the results of such effort. Research on the effort heuristic suggests people judge outcomes as more worthwhile when effort is visible \parencite{kruger2004effort}. As a result, fact-checks of obvious falsehoods may therefore feel more rewarding than corrections of nuanced misinformation. Second, the bridging algorithm's requirement for cross-ideological consensus creates additional evaluation demands beyond simply judging accuracy, as raters must also assess whether opposing partisans will find the note helpful. This meta-cognitive demand may further discourage engagement with nuanced content where cross-partisan agreement seems unlikely.
Third, the incentives to free-ride are substantial, with a large set of experienced raters who are near perfect substitutes in the algorithm. Finally, the absence of any mechanism linking rater compensation or status to engagement with difficult content removes potential incentives that might otherwise offset cognitive effort costs.\footnote{There are some reputational incentives for \emph{writing} community notes, including several third-party ``leader boards'' for prolific helpful note contributors. But no such boards exist for tracking rating contributors. It is also harder to judge whether ratings are productive, since there is no mechanism for rating ratings.} 

Our study addresses a gap in the literature by directly testing whether the reported fact-check difficulty associated with real-world claims predicts Community-Notes success. While prior research has established that crowdsourced fact-checking can be effective when notes appear while identifying various other system vulnerabilities, no work has examined whether the cognitive effort required to evaluate different types of misinformation creates systematic gaps in coverage. By collecting independent ratings of claim believability and difficulty from survey respondents, we can test whether these dimensions predict note publication controlling for content characteristics and partisan dynamics. The results have implications both for understanding the limits of current crowdsourced approaches and for designing systems that might mitigate this difficulty penalty.

\subsection*{Hypotheses}

Drawing on the theoretical framework and established results, above, we derive the following hypotheses:

\begin{itemize}
    \item[\textbf{H1}] \textit{Difficulty $\rightarrow$ Note Failure.} Claims perceived as more difficult to fact-check will be less likely to receive helpful Community Notes. This follows directly from the cognitive miser framework: if raters default to low-effort heuristics, they should selectively disengage from claims requiring substantial evaluation effort.

    \item[\textbf{H2}] \textit{Disengagement Mechanism.} The relationship between difficulty and note failure will operate primarily through rater disengagement (fewer ratings on proposed notes) rather than rater disagreement (lower-quality or more contested ratings). If cognitive laziness drives note failure, we should observe reduced participation rather than increased conflict.

    \item[\textbf{H3}] \textit{Plausibility--Difficulty Link.} Claims perceived as more plausible will also be perceived as more difficult to fact-check. Obviously false claims are easy to refute precisely because their falsity is apparent; plausible misinformation requires deeper investigation.

    \item[\textbf{H4}] \textit{Plausibility $\rightarrow$ Difficulty $\rightarrow$ Note Failure Mediation.} Perceived fact-check difficulty will mediate the relationship between claim plausibility and note failure. That is, plausible misinformation fails to receive notes not simply because it is believable, but because it is believable \textit{and} harder to check, and raters disengage from the resulting cognitive demand.
\end{itemize}

Together, these hypotheses specify a causal chain. We test whether plausible misinformation is harder to fact-check (H3), whether harder claims receive less rater engagement (H2), whether reduced engagement leads to note failure (H1), and whether difficulty mediates the full pathway from plausibility to note failure (H4). If confirmed, this pattern would mean that the cognitive laziness previously documented in individual news consumers \parencite{pennycook2019lazy} also operates at the collective level, creating systematic blind spots in crowdsourced moderation for precisely the misinformation that most requires correction.

\section{Data and Empirical Approach}

To investigate the relationship between fact-check difficulty and the application of Community Notes, we conduct our analysis using the archive of Community Notes data made publicly available by X/Twitter. Our observation window spans between January 1, 2024 and June 30, 2025. Within this period, we identify all posts for which a Community Note had been proposed that contained keywords associated with the COVID-19 vaccine (e.g., ``vaccin*'', ``vaxx*'', ``jab'') and that flagged the post as ``Misinformed or Potentially Misleading''. We further restricted our final dataset to posts that could be successfully retrieved on August 23, 2025, excluding deleted or protected posts to ensure that the content remained visible for subsequent human annotation. This selection process yielded a final corpus of 2,865 unique posts containing vaccine-related claims flagged by Community Notes raters. Following the removal of posts which neither respondent rated as a vaccine claim, this corpus was reduced to 2,250 posts. 

We elected to focus on vaccine-related claims for several reasons, including their potential to cause harm \parencite{loomba2021measuring, pierri2022online} and their prevalence \parencite{mohammadi2025birdwatch, wilson2020social}. Moreover, as vaccine claims are high-salience, morally charged, and consequential, all conditions which should increase the likelihood of motivated engagement \parencite{pennycook2019lazy, kahan2017misconceptions}, we view the focus on vaccine-related claims as a 'least likely' test of the cognitive laziness hypothesis. If effort-sensitive disengagement emerges even in this context, it likely represents a lower bound on bias associated with fact-check difficulty in less salient domains such as finance, climate, or foreign policy. We further examine whether perceived difficulty of fact-checking mediates the relationship between fact-check difficulty and note success. 

For each post in the corpus, we extracted the final adjudication status of the proposed notes on that post. Our primary dependent variable was note publication, encoded as a binary outcome indicating whether any note proposed for a given post achieved the ``Helpful'' status required for public display. To investigate the mechanisms driving note failure, specifically to distinguish between rater disagreement (polarization) and rater neglect (laziness), we further extracted the mix and volume of ratings submitted for each proposed note, as well as metrics of note content, such as word and url counts. Finally, we also collected engagement metrics for each post, specifically the total number of likes and reposts, to serve as controls for the visibility and virality of the content. For some analyses we focus on the ``most-helpful'' note, that which received the most ``helpful'' ratings, as those proposed notes are those that are most likely to actually been applied to the post. This attention to the most-helpful note is justified by the extreme skew of ratings counts. For more than 92\% of our qualifying posts, the most-helpful note received more than 90\% of the total ratings. 

Based on our pre-registered design,\footnote{Found at https://aspredicted.org/aj868h.pdf} we quantified the plausibility and verification difficulty of the claims through a crowd-sourced annotation task recruiting 382 participants via Prolific.\footnote{The research design was evaluated and approved by [anonymized]’s Office of Human Subjects (IRB: \#2025-1270).} Participants were required to be at least 18 years of age and residents of the United States. Each participant evaluated a randomly assigned subset of 15 posts from our corpus, with each post assigned to two participants at random to minimize the influence of individual respondents.\footnote{An error in the integration of images into the Qualtrics survey affected the fifteenth image included in each survey, resulting in their removal from the analysis. Since these images were assigned at random this should not influence the results.} To ensure that ratings reflected the content of the claims rather than the influence of the moderation system, posts were presented in isolation, stripped of any existing Community Notes or status indicators, but retaining all account information which would have been present during the initial rating process so as to not remove critical details. Prior to rating posts participants were asked to determine whether the post made a claim about vaccine safety to ensure rated posts were falsifiable.\footnote{The wording of these central questions are contained in the Appendix.} For each post, participants assessed three dimensions on 5-point Likert scales: personal plausibility (\textit{``How believable is the provided claim?''}), estimates of perceived audience plausibility (\textit{``How believable do you think others would find the claim?''}), and, critically, perceived fact-check difficulty (\textit{``In your opinion, how difficult would it be to fact-check the claim?''}).\footnote{IRR was modest (ICC(1,1) = 0.075 for difficulty, 0.102 for believability), consistent with diverse raters evaluating subjective attributes of heterogeneous stimuli. Because measurement error in an independent variable attenuates effects toward zero, our individual-level estimates are conservative; post-level analyses yield stronger results (see Appendix~\ref{app:robustness}).}

To evaluate whether a relationship exists between fact-check difficulty and community-note application, we estimate a series of regression models relating various elements of the community-notes process to fact-check difficulty ratings. Because each post was rated by two participants, our analysis dataset contains 3,512 individual-level observations (2,250 posts $\times$ 2 raters minus removed data). Rather than averaging rater responses, which would discard within-post variation, we retained the full stacked structure and employed robust standard errors clustered at the post level to account for the non-independence of ratings for the same content. We first investigate whether fact-check difficulty relates to the likelihood that a proposed community-note indicating false or misleading content is actually applied to the message. We estimate both bivariate logit models and models that control for the log-transformed count of likes, reposts, and poster follower counts, to isolate the effect of fact-check difficulty from the confounding effects of content popularity. Next, we also run a series of OLS regressions, controlling for likes, reposts, and followers, relating fact-check difficulty to several other characteristics of the note and the ratings: number of notes, number of ratings and mix of ratings, word- and url-counts in the notes. Finally, we conduct a duration analysis using the Cox proportional-hazards model to evaluate the relationship between fact-check difficult and the length of time that passes between when a post is made and the first community note is applied to it. Figure~\ref{fig:flowchart} in the Appendix provides an overview of the study design, data sources, and analytic pipeline.

\section{Results}

We present our results in four sections. First, we test whether fact-check difficulty predicts note application (H1) and examine the mechanism through which this occurs (H2). Second, we establish the link between plausibility and difficulty (H3) and formally test the mediation pathway (H4). Third, we quantify the real-world consequences of these patterns through counterfactual simulations and time-weighted exposure analysis. Fourth, we present robustness checks addressing alternative explanations.

\subsection{Fact-Check Difficulty Predicts Note Failure (H1)}

\begin{figure}[H]
\centering
\includegraphics[width=\linewidth]{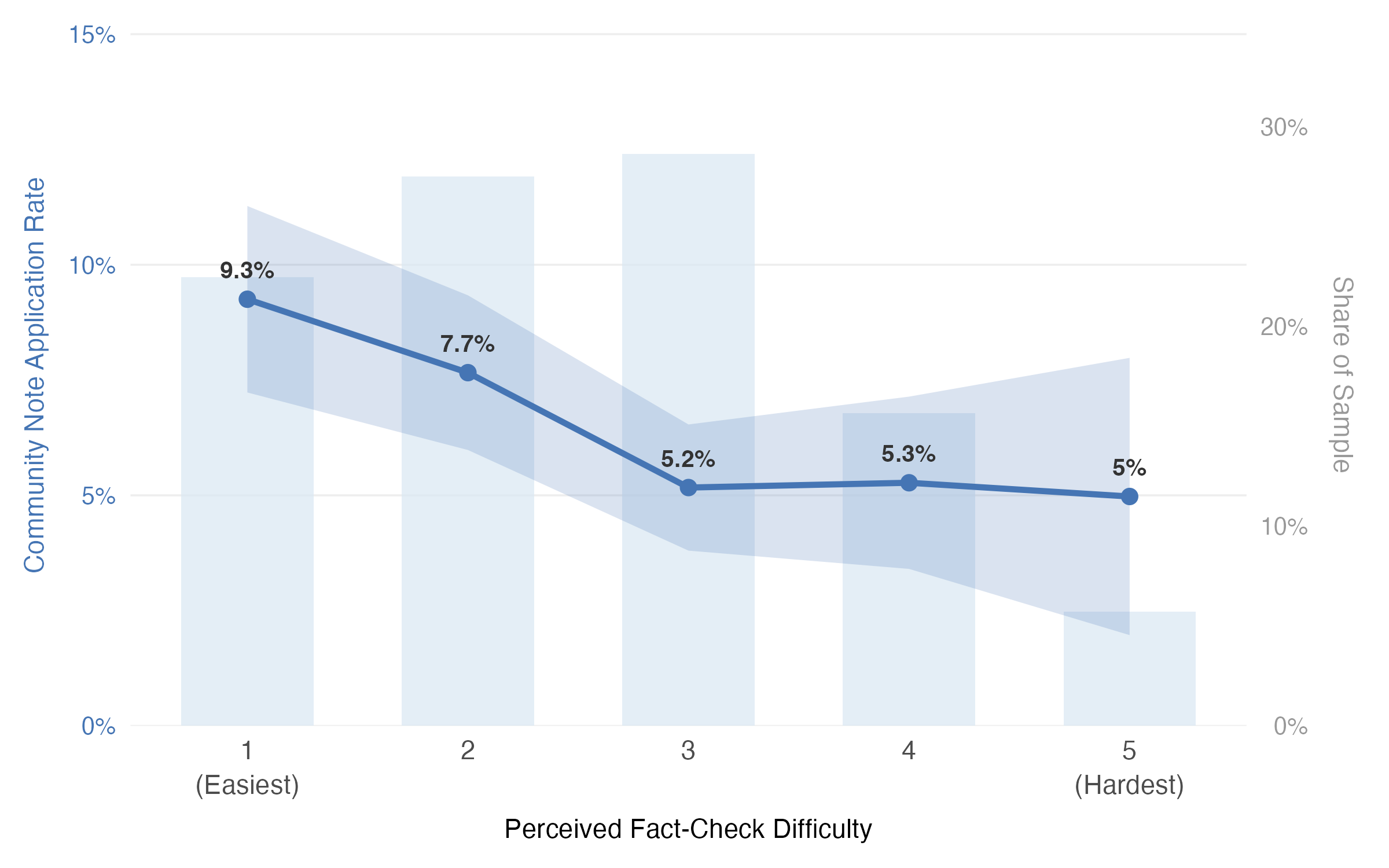}
\caption{Fact-Check Difficulty and Rates of Community Note Application\\
\footnotesize{Note: Solid line represents the share of posts receiving at least one helpful Community Note at each level of perceived fact-check difficulty, with shaded region representing 95\% confidence intervals. Light bars show the share of sample at each difficulty level. $N$ = 3,512 individual judgments across 2,250 posts.}}
\label{fig:difficulty_noterates}
\end{figure}

Figure~\ref{fig:difficulty_noterates} presents both the distribution of fact-check difficulty judgments and how the rate at which community notes are applied varies across those judgments. Application rates decrease monotonically with fact-check difficulty. Posts judged as most difficult to fact-check had only a 5.0\% probability of receiving a note compared to 9.3\% for those judged the easiest to fact-check, a 46\% relative reduction. In a logistic regression predicting community note application with controls for likes, reposts, and followers on the original post, fact-check difficulty significantly predicted note appearance (OR = 0.81, 95\% CI [0.71, 0.91], $p < .001$), confirming H1.

\subsubsection{Disengagement Not Disagreement (H2)}

\begin{figure}[H]
\centering
\includegraphics[width=\linewidth]{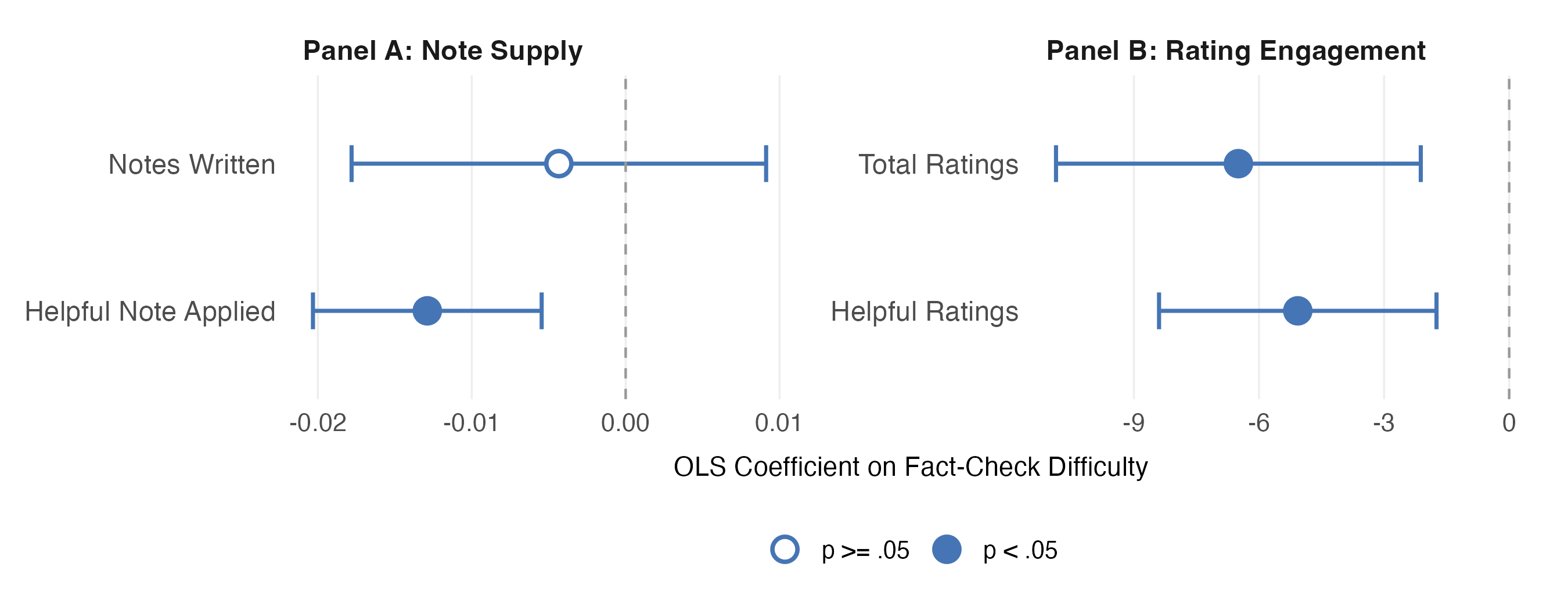}
\caption{Effect of Fact-Check Difficulty on Community Notes Outcomes\\
\footnotesize{Note: OLS coefficients on fact-check difficulty score with robust standard errors clustered at the post level. Controls: log(followers), log(retweets), log(likes). Panel B outcomes measured on the highest-rated note per post. Filled markers indicate $p < .05$; hollow markers indicate $p \geq .05$.}}
\label{fig:reg_coeff}
\end{figure}

A decline in note application could reflect either a supply-side failure (fewer notes written for harder claims) or a demand-side failure in which notes are written but the crowd does not engage with them. Figure~\ref{fig:reg_coeff} distinguishes between these channels.

\textit{Note supply (Panel A).} Fact-check difficulty does not reduce the number of notes written (OLS: $\beta = -0.003$, $p = .70$). Notes are proposed for difficult claims at the same rate as easy ones, yet, as predicted, difficult claims are significantly less likely to receive a helpful note (OLS: $\beta = -0.013$, $p < .001$). The bottleneck is not in note production but in what happens afterward.

\textit{Rating engagement (Panel B).} In addition, notes on harder-to-fact-check posts receive significantly fewer total ratings ($\beta = -6.49$, $p = .004$) and fewer helpful ratings specifically ($\beta = -5.07$, $p < .01$). Critically, the decline in not-helpful ratings is not significant ($\beta = -1.32$, $p = .15$), indicating that raters disengage from difficult claims entirely rather than showing up to disagree. A share-based measure is consistent with the pattern: the proportion of ratings marked ``Helpful'' trends lower for harder claims ($\beta = -0.008$, $p = .065$).

Among rating sub-reasons, notes on harder claims receive a significantly higher share of ratings denoting that the notes were ``Incomplete -- Missing Key Points'' ($\beta = 0.007$, $p < .05$) and a lower share of notes described as ``Addresses the Key Claim'' ($\beta = -0.007$, $p = .12$; see Appendix Table~\ref{tab:subreason}), suggesting that the notes written for difficult claims are themselves perceived as less thorough. Importantly, sub-reasons related to factual accuracy (``Incorrect''), source quality (``Sources Missing or Unreliable''), and bias (``Argumentative or Biased'') show no relationship with difficulty (all $p > .50$; see Appendix Table~\ref{tab:subreason}), indicating that raters did not judge harder claims as more likely to be false.

Taken together, these results support H2. Notes are written for difficult claims, but the crowd does not do the work needed to evaluate them. The failure is one of disengagement (raters simply do not show up) rather than disagreement among those who do participate.

\subsection{Difficulty Mediates the Plausibility Effect (H3, H4)}

The finding that difficult-to-check claims receive fewer notes would not be particularly concerning if all false claims were equally harmful. After all, prioritizing easy-to-check claims might maximize the total number of successful fact-checks. However, if claims that are harder to check also happen to be more plausible, and thus more dangerous, then the pattern documented above represents a systematic tendency to under-address the misinformation that matters most.

\subsubsection{Plausibility Tracks Difficulty (H3)}

Worryingly, posts judged as more difficult to fact-check are also judged as substantially more believable. In bivariate OLS regressions, fact-check difficulty strongly predicted both self-rated believability ($\beta = 0.32$, 95\% CI [0.29, 0.36], $p < .0001$) and perceived audience believability ($\beta = 0.16$, 95\% CI [0.13, 0.20], $p < .0001$). Substantively, posts judged as easiest to fact-check received mean others (self) believability scores of 2.46 (1.70), while those judged most difficult received scores of 3.06 (2.80), increases of 24\% (65\%), confirming H3.

\begin{figure}[H]
\centering
\includegraphics[width=\linewidth]{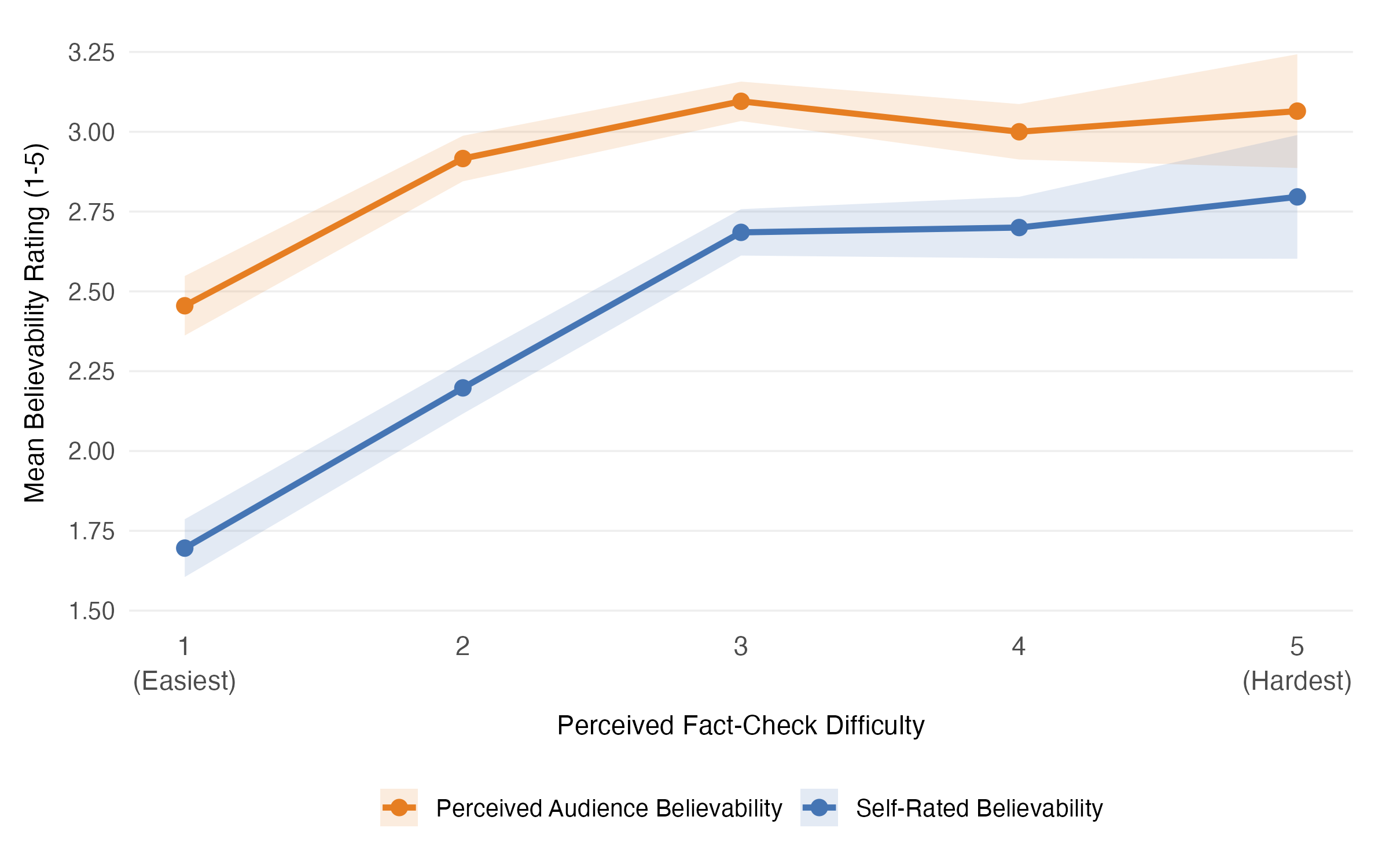}
\caption{Fact-Check Difficulty and Believability\\
\footnotesize{Note: Solid lines represent mean judgments of how believable the underlying post is to the judges themselves (self-rated) and how believable they judge it to be to others (perceived audience). Shaded regions represent 95\% confidence intervals.}}
\label{fig:believe_results}
\end{figure}

\begin{figure}[H]
    \centering
    \includegraphics[width=0.75\textwidth]{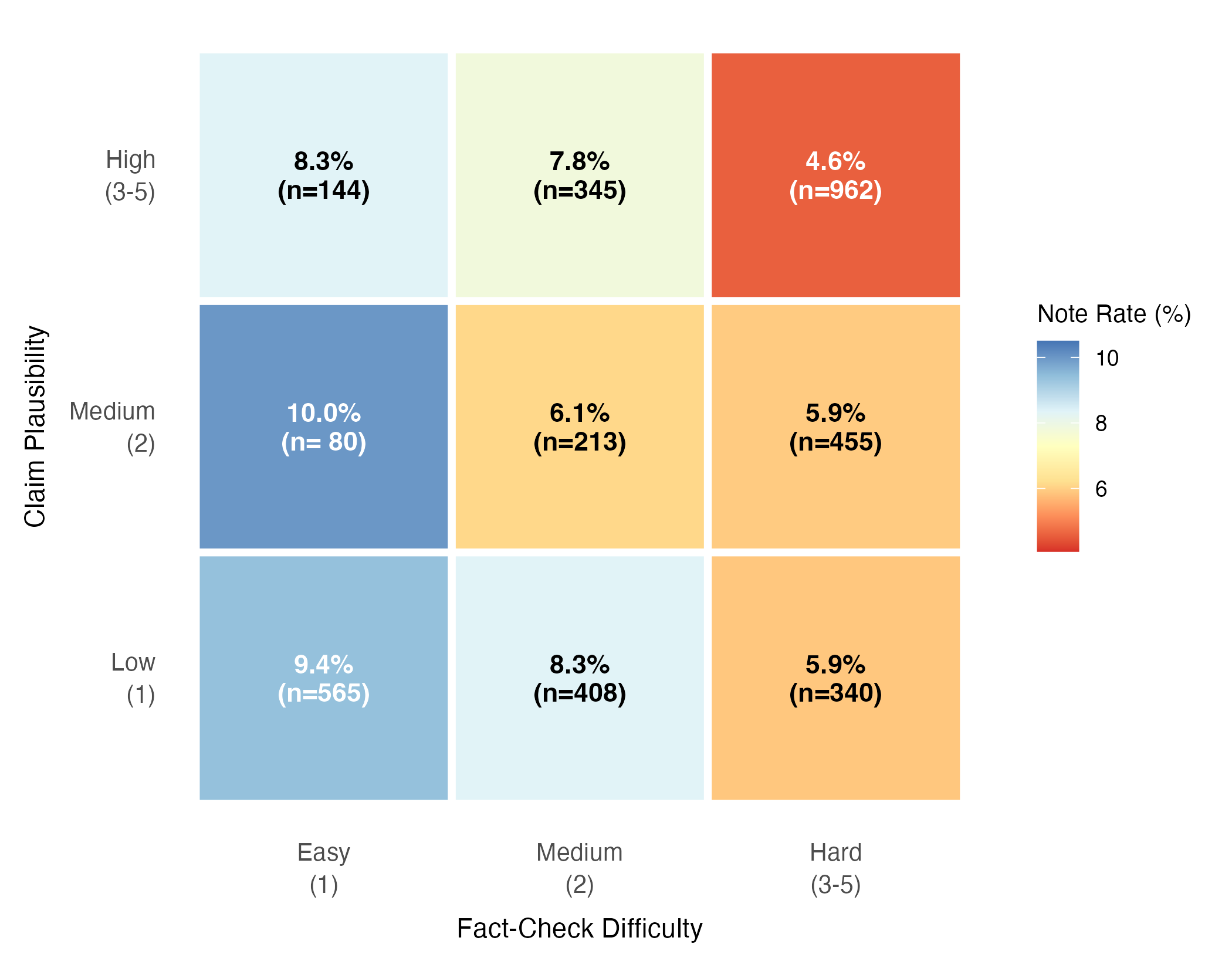}
    \caption{Note Application Rate by Plausibility $\times$ Difficulty.\\
    \footnotesize{Note: Cell area is proportional to sample size ($n$); color indicates note application rate (red = less protection, blue = more protection). The largest cell, which includes high plausibility and hard to fact-check ($n = 962$, 4.6\%), receives notes at less than half the rate of low-plausibility, easy-to-check claims (9.4\%). This pattern holds across alternative binning strategies (see Appendix Figure~\ref{fig:heatmap_binning}).}}
    \label{fig:heatmap}
\end{figure}

To test whether plausibility and difficulty independently predict note failure (rather than simply proxying for one another) we decomposed note application rates in a cross-tabulation of the two dimensions. We grouped individual-level survey responses into three categories of self-rated plausibility (Low: 1, Medium: 2, High: 3--5) and perceived fact-check difficulty (Easy: 1, Medium: 2, Hard: 3--5). The resulting $3 \times 3$ matrix (Figure~\ref{fig:heatmap}) reveals that both dimensions independently predict note failure. Among low-plausibility claims, note application rates decline from 9.4\% for easy-to-check claims to 5.9\% for hard-to-check claims. Among high-plausibility claims, the decline is steeper, dropping from 8.3\% to 4.6\%. Conversely, within the same difficulty level, higher plausibility is associated with lower note rates. The lowest note application rate (4.6\%) occurs at the intersection of high plausibility and high difficulty, which is precisely the category of misinformation most likely to deceive audiences and most in need of correction. Critically, this cell also contains the largest number of observations ($n = 962$), representing over 27\% of the sample.\footnote{Our cross-tabulation results are robust to alternative binning strategies. The gradient from high to low note application rates across plausibility and difficulty holds whether we use 1/2/3--5 bins (main analysis), 1--2/3/4--5 bins, or the full 5-point scale, though cell sizes become small at the extremes of the 5-point scale (minimum $n = 30$). Appendix Figure~\ref{fig:heatmap} presents a side-by-side comparison of both specifications.} 

\subsubsection{Difficulty Mediates the Plausibility Effect (H4)}

To formally test whether perceived fact-check difficulty mediates the relationship between claim plausibility and note application, we conducted a mediation analysis following \parencite{baron1986moderator}'s four-step procedure with a Sobel test for the indirect effect.

All models control for likes, retweets, and followers on the original post. In Step 1 (total effect, $c$ path), we estimated a logistic regression predicting note application from claim plausibility, yielding a marginally significant negative relationship (OR = 0.90, 95\% CI [0.80, 1.01], $p = .063$). In Step 2 ($a$ path), we estimated an OLS regression predicting perceived fact-check difficulty from plausibility, revealing a strong positive association ($b = 0.254$, 95\% CI [0.22, 0.29], $p < .001$): more plausible claims are perceived as substantially harder to fact-check. In Step 3, we entered both plausibility and difficulty as simultaneous predictors of note application. The effect of difficulty remained significant ($b$ path: OR = 0.82, 95\% CI [0.72, 0.94], $p = .004$), while the direct effect of plausibility ($c'$ path) was attenuated and no longer significant (OR = 0.94, 95\% CI [0.84, 1.06], $p = .34$). The Sobel test confirmed that the indirect effect through difficulty was statistically significant (indirect effect = $-0.050$, $Z = -2.86$, $p = .004$), with approximately 48\% of the total effect of plausibility on note failure mediated through perceived fact-check difficulty, confirming H4.

\begin{figure}[htbp]
    \centering
    \includegraphics[width=0.85\textwidth]{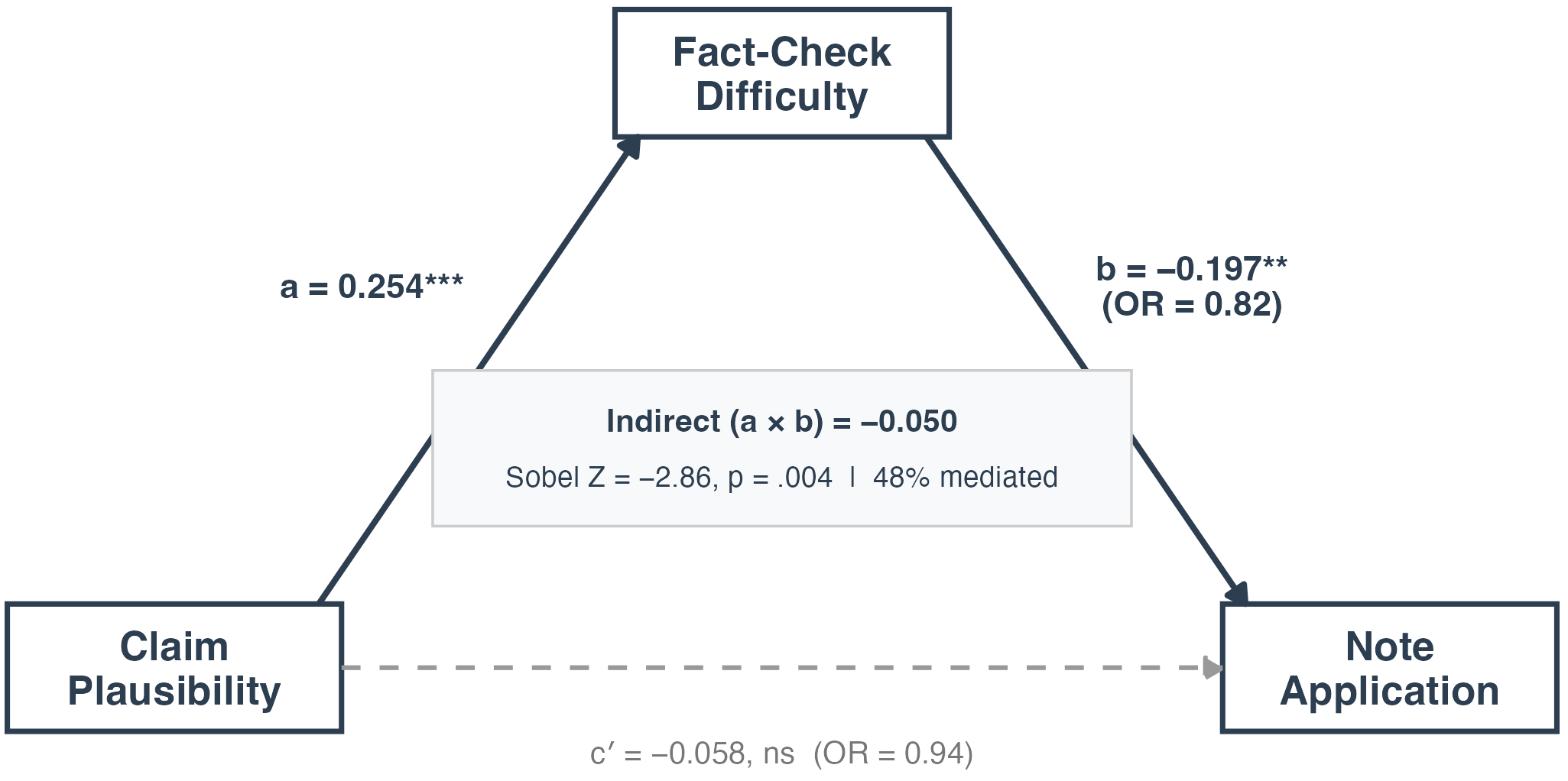}
    \caption{Mediation Analysis: Plausibility $\rightarrow$ Difficulty $\rightarrow$ Note Failure.\\
    \footnotesize{Note: All models control for likes, retweets, and followers on the original post. The $a$ path shows that plausible claims are perceived as harder to fact-check ($b = 0.254$, $p < .001$). The $b$ path shows that difficulty predicts note failure controlling for plausibility (OR = 0.82, $p = .004$). The direct effect of plausibility ($c'$) is attenuated and non-significant when difficulty is included (OR = 0.94, $p = .34$). Sobel test confirms significant mediation ($Z = -2.86$, $p = .004$; 48\% mediated).}}
    \label{fig:mediation}
\end{figure}

These descriptive regularities are consistent with the cognitive laziness account. Specifically, plausible misinformation fails to receive community notes in large part because it is perceived as harder to fact-check, and raters selectively disengage from effortful evaluation tasks. The residual direct effect, though non-significant, suggests that additional mechanisms may contribute, such as raters perceiving less value in correcting plausible claims or feeling less confident in their ability to evaluate them \parencite[the effort heuristic, see][]{kruger2004effort}.

In an extended mediation model, we additionally tested whether difficulty reduces note success through lower rating volume. Perceived difficulty negatively predicted the log-transformed total number of ratings received by the most-helpful note ($b = -0.038$, $p = .054$), and rating volume in turn strongly predicted note application even after controlling for difficulty (OR = 5.24, $p < .001$). While the first stage of this extended pathway is marginally significant, the overall pattern is consistent with selective disengagement. That is, harder-to-check claims attract fewer raters, and fewer raters means fewer opportunities for notes to reach the threshold required for helpful status.


\subsection{Quantifying the Laziness Penalty}

Having established that fact-check difficulty predicts note failure (H1), operates through rater disengagement (H2), and mediates the effect of claim plausibility (H3, H4), we now quantify the real-world consequences of these patterns. 

\begin{figure}[H]
    \centering
    \includegraphics[width=\textwidth]{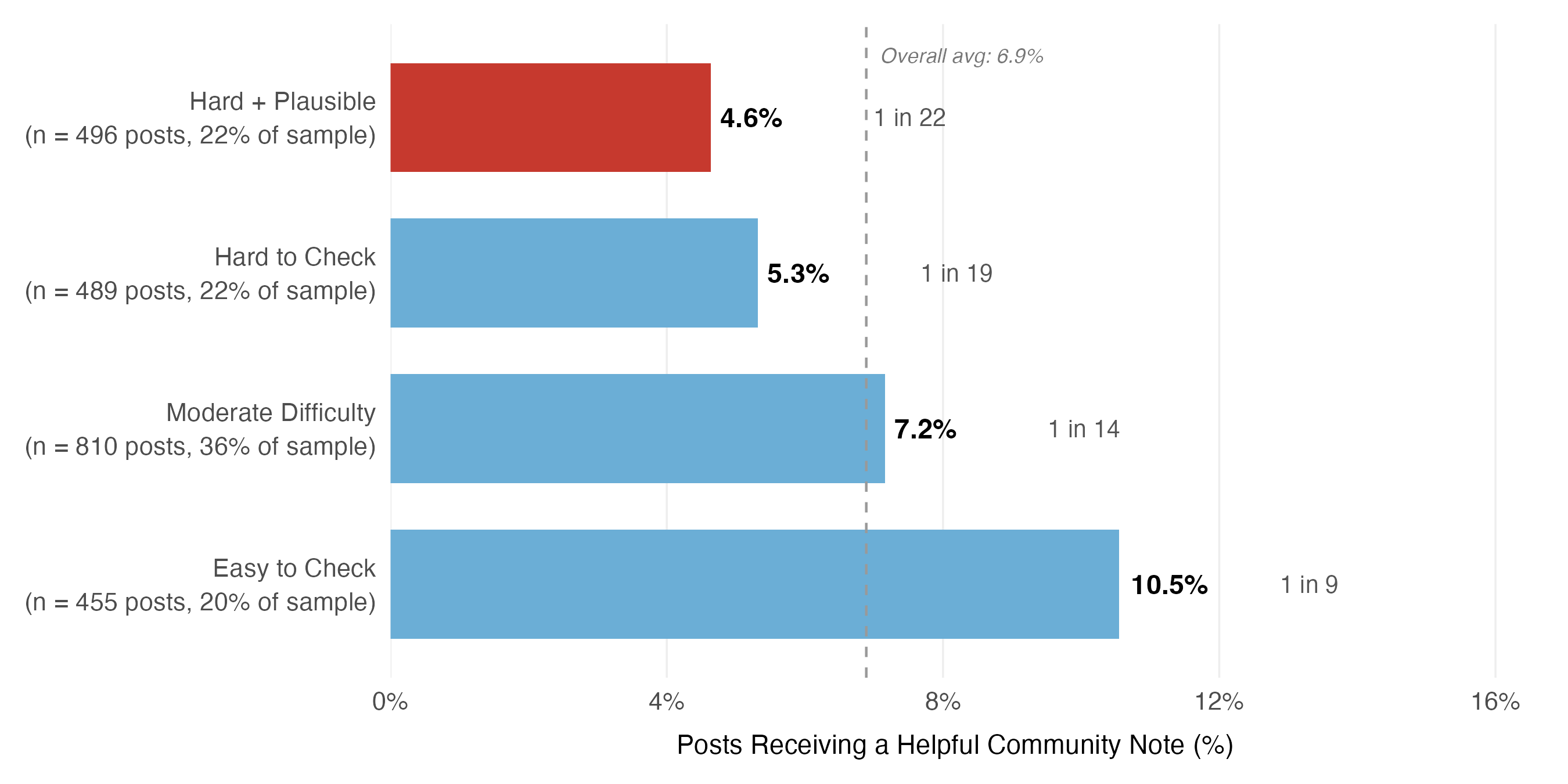}
    \caption{Community Note Correction Rates by Claim Risk Level\\ \footnotesize{Note: Bars show the share of posts receiving at least one helpful Community Note, aggregated at the tweet level (N = 2,250). Categories defined by mean survey ratings: ``Easy to Check'' = fact-check difficulty $<$ 2; ``Moderate Difficulty'' = difficulty 2–3; ``Hard to Check'' = difficulty $\geq$ 3 and self-rated believability $<$ 3; ``Hard + Plausible'' = difficulty $\geq$ 3 and believability $\geq$ 3. Dashed line indicates the overall note application rate (6.9\%). The most dangerous category of misinformation (red) receives corrections at less than half the rate of easily-debunked claims.}}
    \label{fig:consequences}
\end{figure}

Given that only 6.9\% of flagged posts receive helpful notes overall, any systematic bias in which posts receive these rare corrections has outsized consequences. Figure~\ref{fig:consequences} presents note application rates across four categories defined by perceived fact-check difficulty and claim plausibility. The system's correction rate drops by more than half for the content that matters most, ranging from 10.5\% for easily-debunked falsehoods to 4.6\% for claims that are both plausible and difficult to verify (n = 496, 22\% of the sample). Put differently, easy-to-check claims are corrected at roughly 1 in 9, with the most dangerous claims at 1 in 22. A Cox proportional hazards model confirms that harder-to-check claims receive helpful notes at significantly lower rates even after controlling for post engagement (HR = 0.77, 95\% CI [0.65, 0.92], $p < .01$). Among the subset of posts that do receive helpful notes, difficulty does not predict the speed of note application (OLS: $\beta$ = $-$0.08, $p$ = .17), indicating that the barrier operates entirely through the probability channel. As an illustrative calculation, applying \textcite{slaughter2025community}'s recent estimate suggesting that Community Notes reduce reposts by 46\%, eliminating the difficulty penalty would prevent approximately 58,000 additional reposts of hard-to-check potentially harmful misinformation in our sample alone. The crowd's limited fact-checking capacity is thus concentrated on precisely the misinformation least likely to deceive.

\subsection{Ruling Out Alternative Explanations}

Several alternative mechanisms could produce the observed pattern. First, notes on difficult claims might simply be lower quality and thus deserve to fail. Notes on harder-to-check claims contain fewer source URLs ($\beta$ = -0.20, p $<$ .01), and the share of ratings marked ``helpful'' trends in the same direction ($\beta$ = -0.008, p = .065, see Appendix~\ref{app:regressions} and~\ref{app:notequality}). These quality differences are consistent with the effort-aversion account. Claims that are harder to verify are also harder to write compelling notes about and harder to find authoritative sources for. Difficulty thus depresses Community Notes effectiveness through multiple channels. Fewer contributors engage and those who do produce modestly weaker output. However, the quantity margin remains dominant. Substantively, moving from the easiest to the hardest claims in our sample, total ratings on the best note fall by nearly a full 25 ratings, while word count drops by only two words and URL count shifts by less than one link per unit of difficulty. Second, differential post engagement cannot explain the pattern, as all models control for likes, reposts, and follower counts. Third, the laziness penalty might emerge only for claims that are both plausible and difficult rather than operating additively. A logistic regression including a plausibility $\times$ difficulty interaction term, controlling for engagement on the original post, found no significant interaction (OR = 0.97, p = .51), suggesting each dimension independently contributes to note failure.

Fourth, the difficulty penalty might reflect claim veracity rather than verification effort. That is, if harder-to-check claims are actually \textit{true}, raters' reluctance to engage could represent appropriate restraint rather than laziness. To test this, consistent with emerging evidence that LLMs can serve to classify claim veracity \parencite{hoes2023leveraging}, we submitted all 2,250 posts to an automated fact-checking pipeline using GPT-4.1 with web search access and GPT-5 fallback for low-confidence verdicts.\footnote{GPT-5 served as an escalation step: claims receiving low-confidence or ``uncertain'' verdicts from GPT-4.1 were re-evaluated by GPT-5. The primary fact-checking, claim extraction, and timing measurements all relied on GPT-4.1.} See Appendix~\ref{app:llm} for details. GPT does not judge harder-to-check posts as more likely to be true (OR = 1.08, $p = .17$). At the same time, posts that survey respondents rated as harder to check took GPT significantly longer to process (OLS on log-transformed seconds: $\beta = 0.042$, $p = .041$; descriptively, 77 vs. 66 seconds for the hardest and easiest terciles), providing additional validation that perceived difficulty reflects genuine verification complexity. GPT is also significantly less confident in its verdicts for harder posts ($\beta = -0.006$, $p = .033$), consistent with these claims being more ambiguous to adjudicate. In sum, the difficulty construct captures real cognitive demand rather than differences in underlying truth.

\section{Discussion}

Our findings reveal a previously unexamined vulnerability in crowdsourced fact-checking systems: the laziness of the crowd. Claims that are more difficult to fact-check are significantly less likely to receive successful Community Notes, and the disparity is largest for the content that poses the greatest risk. Among posts rated as both highly plausible and hard to verify, only 1 in 22 receives a helpful note, compared with 1 in 9 for easily checked claims, a 56\% relative reduction in the correction rate (Figure~\ref{fig:consequences}). Given that fewer than 8\% of posts in our sample received helpful notes overall, this gap represents a near-halving of an already rare outcome. 

These findings extend the cognitive miser framework \parencite{fiske1991social, stanovich2018miserliness} from individual news consumption to collective fact-checking behavior. \textcite{pennycook2019lazy} established that individuals default to low-effort heuristics when evaluating misinformation, accepting plausible falsehoods because effortful verification feels unnecessary. We demonstrate that this same tendency operates at the level of the collective, but with a structural twist. Unlike canonical wisdom-of-the-crowds settings where participants must respond to a fixed set of stimuli (Galton's fair-goers all estimated the weight of the same ox), Community Notes raters face no such constraint. They can decline to evaluate any claim that seems to require additional effort. Voluntary participation thus transforms collective intelligence into selective intelligence, systematically biased toward tractable problems. The result is not that crowds lack wisdom but that their wisdom is unevenly applied, concentrated on obvious falsehoods while neglecting the plausible misinformation that most requires correction. This pattern is compounded by what \textcite{jalbert2026misaligned} identify as a systematic misalignment between lay moderation intuitions and moderation effectiveness, as survey participants preferentially elect to remove implausible content, despite evidence that only removing plausible content reduced broader conspiratorial beliefs.

These results complement recent work on Community Notes effectiveness and vulnerabilities \textcite{vraga2025understanding}. \textcite{slaughter2025community} demonstrated that notes reduce engagement with misinformation by approximately 46\% when they appear, but noted a substantial drop off in efficacy after just a half-day delay. Our findings suggest one reason for these damaging delays: plausible misinformation may languish at the ``Needs More Ratings'' status because raters are unwilling to invest the cognitive effort required to evaluate complex claims. The crowd is effective when it acts, but `laziness' may prevent members of the crowd from acting on the claims that matter most.

Similarly, \textcite{truong2025community} identified algorithmic vulnerabilities in Community Notes, demonstrating that the bridging algorithm suppresses over 40\% of genuinely helpful notes and remains sensitive to coordinated manipulation. Our findings add a complementary concern by illustrating that even absent manipulation, the system may fail organically where raters selectively engage with tractable content. The combination of algorithmic suppression and rater laziness suggests that crowdsourced moderation faces challenges on multiple fronts.

These findings point toward the need for several design interventions. First, platforms should implement hybrid routing systems that direct high-plausibility, low-engagement claims to professional fact-checkers. Machine learning classifiers could estimate claim plausibility at submission time and automatically escalate claims that fail to attract sufficient ratings within a defined window, leveraging the crowd's capacity for high-volume processing of straightforward cases while reserving professional resources for claims where volunteer effort is insufficient \parencite{ho2013adaptive, dai2013pomdp}. Second, platforms should introduce differential incentives that reward engagement with difficult claims. The current reputation system awards points equally regardless of task difficulty, creating no incentive to tackle hard problems. Implementing difficulty-weighted reputation gains, where evaluating notes on high-plausibility claims earns greater status, could offset the cognitive costs that discourage engagement with nuanced misinformation \parencite{singer2013pricing}.

Third, given that timeliness determines correction effectiveness \parencite{slaughter2025community}, platforms should (and have begun to for certain types of notes) implement automated nudges that surface stalled notes to high-reputation raters. When notes on high-engagement posts remain in ``Needs More Ratings'' status beyond a threshold period, the system could send targeted notifications to raters with demonstrated expertise in the relevant domain. None of these solutions is straightforward. Difficulty-weighted incentives require platforms to correctly identify task difficulty, and hybrid routing partially undermines the voluntary model's scalability. At the same time, all are more tractable than correcting for biases that platforms cannot directly observe.

These findings suggest a fundamental tension in the design of voluntary content moderation systems. Crowdsourcing achieves scale by distributing effort across many contributors, but voluntary participation means that contributors can optimize their own effort allocation. When individual optimization leads contributors to avoid cognitively demanding tasks, the system inherits a collective bias toward tractable problems regardless of their importance. This represents a distinct failure mode from those previously identified in Community Notes research, as it relates to neither algorithmic suppression \parencite{truong2025community} nor partisan disagreement \parencite{allen2022birds}, but rather systematic inattention. Correcting this bias requires either eliminating task selection (which would undermine the voluntary model's scalability) or restructuring incentives to make difficult tasks more attractive (which requires platforms to correctly identify and weight task difficulty). Neither solution is straightforward, but both are more tractable than correcting for biases that platforms cannot observe.


Several limitations temper our confidence and warrant acknowledgment. First, our plausibility and difficulty ratings were collected from survey participants rather than actual Community Notes raters, who may differ systematically in motivation and expertise. Second, we intentionally focused on vaccine-related claims as a stress test for the cognitive laziness hypothesis. As this topic is highly salient and emotionally charged, it represents a case where raters should be most motivated to overcome cognitive inertia. The fact that difficulty penalty remains a significant predictor of note failure in this high-stakes environment suggests that the bias is likely even more pronounced in less contentious domains. Future research should apply this framework to areas like financial fraud or emerging political narratives where the incentives for effortful verification may be lower. Third, our design cannot fully rule out alternative explanations such as differential initial note responsiveness to claims of varying plausibility, though our analyses suggest this is unlikely to fully account for the observed patterns. Future work should examine whether the identified difficulty penalty operates similarly across content domains and whether the proposed interventions successfully mitigate it.

\section*{Conclusion}

Crowdsourced moderation has emerged as a promising response to the scalability constraints facing professional fact-checkers, and our findings temper this optimism by identifying a systematic blind spot in the ``crowd's'' unwillingness to invest equivalent cognitive effort in evaluating complex claims. The difficulty penalty is not trivial. The most dangerous misinformation in our sample, claims that are both plausible and hard to verify, receives corrective notes at less than half the rate of easily checked falsehoods, with only 1 in 22 such posts receiving a helpful note. Community Notes and similar systems may be well-suited for flagging obvious falsehoods but ill-equipped to address the forms of plausible misinformation that pose the greatest threat to informed public discourse. With crowdsourced fact-checking becoming the main system for policing falsehoods at several other large platforms, including YouTube, Meta, and TikTok, addressing this limitation will require platform designs that acknowledge human cognitive constraints and possibly greater investment in nascent hybrid systems that route difficult content to experts, incentive structures that reward effortful engagement, and automated interventions that overcome the crowd's inertia. Without such reforms, crowdsourced moderation risks becoming a tool that corrects only the misinformation least likely to deceive.

\clearpage
\printbibliography
\clearpage

\appendix
\renewcommand{\thetable}{A\arabic{table}}
\renewcommand{\thefigure}{A\arabic{figure}}
\setcounter{table}{0}
\setcounter{figure}{0}

\noindent\textbf{Appendix Contents}
\vspace{4pt}

\noindent\begin{tabular}{@{}l l r@{}}
\toprule
& Section & Page \\
\midrule
\ref{app:studydesign}  & \nameref{app:studydesign}   & \pageref{app:studydesign} \\
\ref{app:sample}       & \nameref{app:sample}        & \pageref{app:sample} \\
\ref{app:descriptives} & \nameref{app:descriptives}   & \pageref{app:descriptives} \\
\ref{app:survey}       & \nameref{app:survey}         & \pageref{app:survey} \\
\ref{app:construct}    & \nameref{app:construct}      & \pageref{app:construct} \\
\ref{app:regressions}  & \nameref{app:regressions}    & \pageref{app:regressions} \\
\ref{app:notequality}  & \nameref{app:notequality}    & \pageref{app:notequality} \\
\ref{app:textfeatures} & \nameref{app:textfeatures}   & \pageref{app:textfeatures} \\
\ref{app:cox}          & \nameref{app:cox}            & \pageref{app:cox} \\
\ref{app:robustness}   & \nameref{app:robustness}     & \pageref{app:robustness} \\
\ref{app:binning}      & \nameref{app:binning}        & \pageref{app:binning} \\
\ref{app:llm}          & \nameref{app:llm}            & \pageref{app:llm} \\
\ref{app:instructions}          & \nameref{app:instructions}            & \pageref{app:instructions} \\
\bottomrule
\end{tabular}

\vspace{12pt}

\clearpage
\section{Study Design}\label{app:studydesign}

\begin{figure}[H]
\centering
\begin{tikzpicture}[
    node distance=1cm and 1cm,
    box/.style={rectangle, draw, minimum width=3.2cm, minimum height=0.8cm, align=center, font=\small},
    widebox/.style={rectangle, draw, thick, minimum width=4.5cm, minimum height=0.8cm, align=center, font=\small},
    smallbox/.style={rectangle, draw, minimum width=2.6cm, minimum height=0.7cm, align=center, font=\footnotesize},
    arrow/.style={->, >=stealth, thick}
]

\node[box] (cn) {Community Notes Archive\\{\footnotesize 2,250 posts}};
\node[box, right=1cm of cn] (survey) {Prolific Survey\\{\footnotesize 382 participants}};
\node[box, right=1cm of survey] (llm) {LLM Validation\\{\footnotesize GPT-4.1 + GPT-5}};

\node[widebox, below=1.5cm of survey] (merge) {\textbf{Merged Dataset}\\{\footnotesize $N$ = 3,512; clustered SEs}};

\draw[arrow] (cn) -- (merge);
\draw[arrow] (survey) -- (merge);
\draw[arrow, dashed] (llm) -- (merge);

\node[smallbox, below left=1.2cm and 1.5cm of merge] (h12) {\textbf{H1/H2}\\{\scriptsize Difficulty $\to$ Note failure}};
\node[smallbox, below=1.2cm of merge] (h34) {\textbf{H3/H4}\\{\scriptsize Mediation via difficulty}};
\node[smallbox, below right=1.2cm and 1.5cm of merge] (cox) {\textbf{Cox PH}\\{\scriptsize Time to helpful note}};

\draw[arrow] (merge) -- (h12);
\draw[arrow] (merge) -- (h34);
\draw[arrow] (merge) -- (cox);

\end{tikzpicture}
\caption{\textit{Study design. Community Notes archive data and Prolific survey ratings are merged at the post level. LLM fact-checking (dashed) serves as external validation.}}
\label{fig:flowchart}
\end{figure}
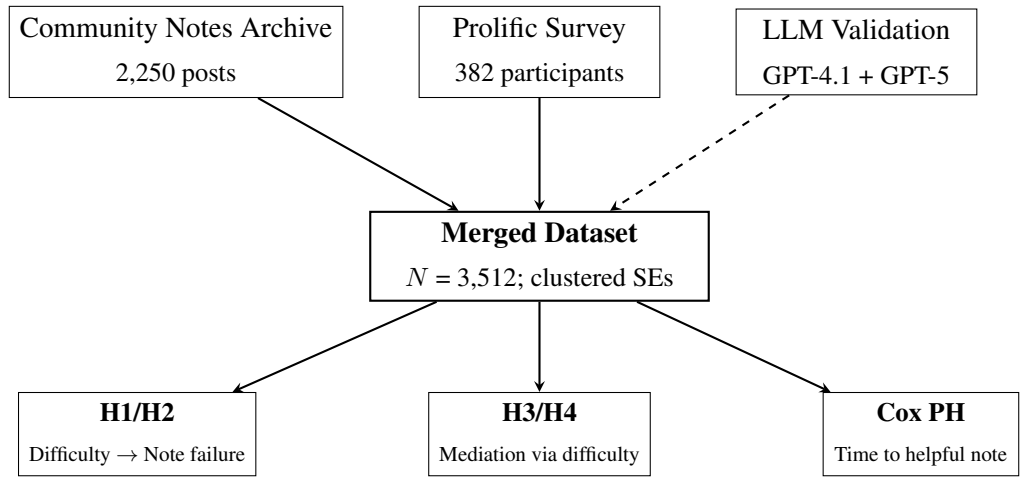

\clearpage
\section{Sample Characteristics}\label{app:sample}

Table~\ref{tab:demographics} presents the demographic characteristics of our survey sample. Participants were recruited via Prolific and were required to be at least 18 years of age and residents of the United States.

\begin{table}[H]
\centering
\caption{Demographic Characteristics of Survey Participants ($N = 382$)}
\label{tab:demographics}
\begin{tabular}{lrr}
\toprule
Characteristic & $N$ & \% \\
\midrule
\textbf{Gender} \\
\quad Female & 194 & 50.8 \\
\quad Male & 187 & 49.0 \\
\midrule
\textbf{Ethnicity} \\
\quad White & 256 & 67.0 \\
\quad Black & 47 & 12.3 \\
\quad Mixed & 39 & 10.2 \\
\quad Asian & 18 & 4.7 \\
\quad Other & 21 & 5.5 \\
\midrule
\textbf{Age} \\
\quad 18--29 & 64 & 16.8 \\
\quad 30--39 & 82 & 21.5 \\
\quad 40--49 & 59 & 15.4 \\
\quad 50--59 & 52 & 13.6 \\
\quad 60--69 & 76 & 19.9 \\
\quad 70+ & 48 & 12.6 \\
\bottomrule
\end{tabular}
\end{table}

\clearpage
\section{Descriptive Statistics}\label{app:descriptives}

Table~\ref{tab:survey_dist} presents the distribution of survey responses for our three key measures. Each post was rated by two participants. An error in the survey platform affected one image per participant, resulting in 384 missing observations (9.7\% of the sample). Since affected images were randomly assigned, this missingness should not bias results.

\begin{table}[H]
\centering
\caption{Distribution of Survey Responses}
\label{tab:survey_dist}
\begin{tabular}{lrr}
\toprule
Measure & $N$ & \% \\
\midrule
\textbf{Personal Believability} ($N = 3{,}566$) \\
\quad 1 -- Not at all believable & 1,338 & 37.5 \\
\quad 2 -- Slightly believable & 760 & 21.3 \\
\quad 3 -- Moderately believable & 663 & 18.6 \\
\quad 4 -- Believable & 551 & 15.5 \\
\quad 5 -- Very believable & 254 & 7.1 \\
\midrule
\textbf{Perceived Audience Believability} ($N = 3{,}566$) \\
\quad 1 -- Not at all believable & 468 & 13.1 \\
\quad 2 -- Slightly believable & 919 & 25.8 \\
\quad 3 -- Moderately believable & 1,077 & 30.2 \\
\quad 4 -- Believable & 765 & 21.5 \\
\quad 5 -- Very believable & 337 & 9.4 \\
\midrule
\textbf{Fact-Check Difficulty} ($N = 3{,}566$) \\
\quad 1 -- Very easy & 809 & 22.7 \\
\quad 2 -- Easy & 980 & 27.5 \\
\quad 3 -- Moderate & 1,022 & 28.7 \\
\quad 4 -- Difficult & 552 & 15.5 \\
\quad 5 -- Very difficult & 203 & 5.7 \\
\bottomrule
\end{tabular}
\end{table}

Table~\ref{tab:cn_outcomes} presents the distribution of Community Notes outcomes for posts in our sample. The analytic sample comprises all 2,250 posts rated by survey participants, all of which were successfully matched to engagement metadata (followers, retweets, likes) used as controls.

\begin{table}[H]
\centering
\caption{Community Notes Outcomes ($N = 2{,}250$ posts in analytic sample)}
\label{tab:cn_outcomes}
\begin{tabular}{lrr}
\toprule
Outcome & $N$ & \% \\
\midrule
\textbf{Helpful Note Applied} \\
\quad Yes & 155 & 6.9 \\
\quad No & 2,095 & 93.1 \\
\midrule
\textbf{Notes Proposed per Post} \\
\quad Median & \multicolumn{2}{c}{1.0} \\
\quad Mean & \multicolumn{2}{c}{1.15} \\
\quad Range & \multicolumn{2}{c}{1--8} \\
\bottomrule
\end{tabular}
\end{table}

\clearpage
\section{Survey Instrument}\label{app:survey}

\subsection{Example Stimulus}
\begin{figure}[H]
\centering
\includegraphics[width=0.5\linewidth]{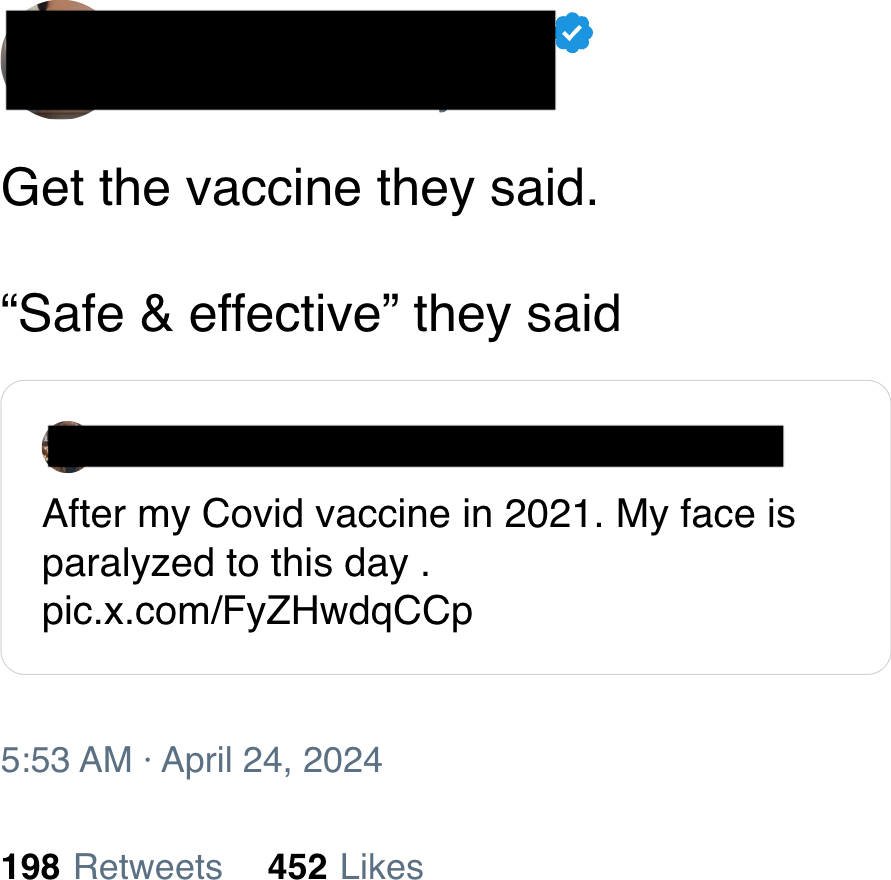}
\caption{Example of Post Stimulus as Presented to Participants}
\label{fig:stimulus}
\end{figure}

Figure~\ref{fig:stimulus} shows an example post as presented to participants. All posts were displayed as screenshot images with account names redacted.

\clearpage

\section{Construct Validity of Perceived Fact-Check Difficulty}\label{app:construct}

A natural concern is that perceived fact-check difficulty does not represent a distinct construct but instead proxies for believability, topic complexity, or rhetorical style. We address this in three ways.

\subsection{Discriminant Validity}

Fact-check difficulty is only modestly correlated with personal believability ($r = .29$) and weakly correlated with perceived audience believability ($r = .17$). Self-rated believability explains just 8.2\% of the variance in difficulty ratings, leaving over 90\% unique to the difficulty construct. In a multiple regression predicting difficulty from both believability measures, self-rated believability is strongly predictive ($\beta = 0.265$, $p < .001$) but perceived audience believability adds no incremental prediction ($\beta = -0.019$, $p = .35$). The partial correlation between difficulty and audience believability, controlling for self-rated believability, is effectively zero ($r_{\text{partial}} = -.02$). This suggests that the shared variance between difficulty and believability runs through a personal assessment of the claim's plausibility rather than a social judgment about what others might believe.

\subsection{Linguistic Properties}

We extracted 15 surface-level linguistic features from each tweet's text and regressed perceived difficulty on all features simultaneously (see Table~\ref{tab:text_features} in Appendix~\ref{app:textfeatures}). Sensationalist markers (exclamation points, ALL CAPS, emoji), definitive language (``proven,'' ``confirmed,'' ``BREAKING''), references to named organizations (FDA, WHO, Pfizer), and appeals to authority are all unrelated to perceived difficulty. Only word count reaches significance ($\beta = 0.005$, $p = .001$), while hedge language (words like ``may,'' ``could,'' ``reportedly,'' and ``suggests'') trends positive but does not reach significance ($\beta = 0.133$, $p = .065$). Collectively, all measured text features explain just 1.1\% of the variance in difficulty ratings, indicating that perceived difficulty reflects a deeper assessment of claim content rather than rhetorical style. The hedge-language trend is nonetheless theoretically informative: claims framed as possibilities rather than certainties tend to be perceived as harder to fact-check, consistent with the intuition that hedged claims resist simple falsification because they do not make clear, testable assertions.

\subsection{Community Notes Rater Behavior}

If perceived difficulty captures something real about the verifiability of a claim, it should manifest in how Community Notes raters themselves evaluate proposed notes. We tested this by regressing each sub-reason share (from raters' helpfulness assessments of the top note per post) on perceived difficulty. As difficulty increases, raters are significantly more likely to cite ``Missing Key Points'' as a reason for marking the note unhelpful ($\beta = 0.008$, $p = .026$), and marginally less likely to endorse ``Addresses the Claim'' as a reason for marking it helpful ($\beta = -0.007$, $p = .097$). Crucially, reasons related to factual accuracy (``Incorrect''; $\beta = 0.002$, $p = .51$), source quality (``Sources Missing or Unreliable''; $\beta = 0.000$, $p = .93$), and bias (``Argumentative or Biased''; $\beta = 0.000$, $p = .94$) show no relationship with difficulty. This pattern suggests that what makes a claim difficult to fact-check is not that evidence against it is unavailable, but that it is \textit{incomplete}. That is, the claim resists the kind of clean, comprehensive rebuttal that earns high marks from the community.

\subsection{Summary}

Perceived fact-check difficulty is a distinct construct from believability, is not driven by surface-level textual features, and manifests in the Community Notes ecosystem as a problem of epistemic completeness. Difficult claims elicit notes that raters perceive as failing to fully address the core assertion, even when the notes are factually accurate and well-sourced. This supports our theoretical interpretation that difficulty captures genuine cognitive demand rather than serving as a proxy for other claim characteristics.

\clearpage
\section{Full Regression Tables}\label{app:regressions}

Tables~\ref{tab:panel_a}--\ref{tab:subreason} report the full OLS regression results corresponding to Figure~2 in the main text. All models use robust standard errors clustered at the post level. Controls are log(followers), log(retweets), and log(likes).

\begin{table}[H]
\centering
\caption{OLS Regressions: Note Supply Outcomes}
\label{tab:panel_a}
\begin{tabular}{lcccc}
\toprule
& \multicolumn{2}{c}{Notes Written} & \multicolumn{2}{c}{Helpful Note Applied} \\
\cmidrule(lr){2-3} \cmidrule(lr){4-5}
& (1) & (2) & (3) & (4) \\
\midrule
Fact-Check Difficulty & $-$0.003 & $-$0.004 & $-$0.013$^{***}$ & $-$0.013$^{***}$ \\
& (0.007) & (0.007) & (0.004) & (0.004) \\
log(Likes) & & 0.012 & & 0.007 \\
& & (0.012) & & (0.009) \\
log(Retweets) & & 0.029$^{*}$ & & $-$0.015 \\
& & (0.013) & & (0.009) \\
log(Followers) & & 0.006 & & 0.018$^{***}$ \\
& & (0.005) & & (0.004) \\
Intercept & 1.161 & 0.833 & 0.102 & $-$0.073 \\
& (0.021) & (0.054) & (0.012) & (0.027) \\
\midrule
$N$ & 3,512 & 3,512 & 3,512 & 3,512 \\
Clusters & 2,250 & 2,250 & 2,250 & 2,250 \\
$R^{2}$ & 0.000 & 0.045 & 0.004 & 0.018 \\
DV Mean & 1.154 & 1.154 & 0.068 & 0.068 \\
\bottomrule
\end{tabular}\\
\footnotesize{Robust standard errors clustered at the post level in parentheses. $^{*}p < .05$; $^{**}p < .01$; $^{***}p < .001$.}
\end{table}

\begin{table}[H]
\centering
\caption{OLS Regressions: Rating Engagement Outcomes (Top Note per Post)}\label{tab:panel_b}
\begin{tabular}{lcccccc}
\toprule
& \multicolumn{2}{c}{Total Ratings} & \multicolumn{2}{c}{Helpful Ratings} & \multicolumn{2}{c}{Not-Helpful Ratings} \\
\cmidrule(lr){2-3} \cmidrule(lr){4-5} \cmidrule(lr){6-7}
& (1) & (2) & (3) & (4) & (5) & (6) \\
\midrule
Fact-Check Difficulty & $-$6.69$^{**}$ & $-$6.49$^{**}$ & $-$5.26$^{**}$ & $-$5.07$^{**}$ & $-$1.32 & $-$1.32 \\
& (2.41) & (2.23) & (1.75) & (1.70) & (1.03) & (0.92) \\
log(Likes) & & 51.56$^{***}$ & & 27.63$^{***}$ & & 23.07$^{***}$ \\
& & (8.26) & & (5.01) & & (4.56) \\
log(Retweets) & & $-$28.50$^{***}$ & & $-$18.89$^{***}$ & & $-$9.23$^{*}$ \\
& & (6.79) & & (4.16) & & (3.76) \\
log(Followers) & & 2.96 & & 2.08 & & 0.78 \\
& & (2.84) & & (1.61) & & (1.69) \\
\midrule
$N$ & 2,505 & 2,505 & 2,505 & 2,505 & 2,505 & 2,505 \\
Clusters & 1,538 & 1,538 & 1,538 & 1,538 & 1,538 & 1,538 \\
$R^{2}$ & 0.003 & 0.135 & 0.004 & 0.063 & 0.000 & 0.157 \\
DV Mean & 69.07 & 69.07 & 37.02 & 37.02 & 30.35 & 30.35 \\
\bottomrule
\end{tabular}\\
\footnotesize{Robust standard errors clustered at the post level in parentheses. Sample restricted to posts with at least one note that received ratings. $^{*}p < .05$; $^{**}p < .01$; $^{***}p < .001$.}
\end{table}

\begin{table}[H]
\centering
\caption{OLS Regressions: Rating Composition (Top Note per Post)}
\label{tab:subreason}
\begin{tabular}{lcccccc}
\toprule
& \multicolumn{2}{c}{Share Helpful} & \multicolumn{2}{c}{Addresses Claim} & \multicolumn{2}{c}{Missing Key Points} \\
\cmidrule(lr){2-3} \cmidrule(lr){4-5} \cmidrule(lr){6-7}
& (1) & (2) & (3) & (4) & (5) & (6) \\
\midrule
Fact-Check Difficulty & $-$0.010$^{*}$ & $-$0.008 & $-$0.007 & $-$0.007 & 0.008$^{*}$ & 0.007$^{*}$ \\
& (0.005) & (0.005) & (0.004) & (0.004) & (0.003) & (0.003) \\
log(Likes) & & $-$0.008 & & $-$0.021 & & $-$0.012 \\
& & (0.014) & & (0.013) & & (0.010) \\
log(Retweets) & & $-$0.039$^{**}$ & & $-$0.018 & & 0.031$^{***}$ \\
& & (0.013) & & (0.012) & & (0.009) \\
log(Followers) & & 0.019$^{***}$ & & 0.010$^{*}$ & & $-$0.008$^{*}$ \\
& & (0.005) & & (0.005) & & (0.003) \\
\midrule
$N$ & 2,505 & 2,505 & 2,505 & 2,505 & 2,505 & 2,505 \\
Clusters & 1,538 & 1,538 & 1,538 & 1,538 & 1,538 & 1,538 \\
$R^{2}$ & 0.002 & 0.090 & 0.001 & 0.075 & 0.002 & 0.035 \\
DV Mean & 0.551 & 0.551 & 0.452 & 0.452 & 0.218 & 0.218 \\
\bottomrule
\end{tabular}\\
\footnotesize{Outcomes are shares of total ratings on the most-helpful note per post. Robust standard errors clustered at the post level in parentheses. $^{*}p < .05$; $^{**}p < .01$; $^{***}p < .001$.}
\end{table}

\clearpage
\section{Note Quality by Fact-Check Difficulty}\label{app:notequality}

To assess whether notes written for harder-to-check claims are lower quality, we extracted the text of each Community Note matched to our sample and computed word count and URL count for the highest-rated note per post. Table~\ref{tab:note_quality_desc} reports descriptive means by difficulty tercile, and Table~\ref{tab:note_quality_ols} reports OLS regressions of note quality on perceived fact-check difficulty.

\begin{table}[H]
\centering
\caption{Note Quality Descriptives by Fact-Check Difficulty}
\label{tab:note_quality_desc}
\begin{tabular}{lcccc}
\toprule
& Easy (1--2) & Medium (3) & Hard (4--5) & Overall \\
\midrule
Word Count & 39.0 & 38.0 & 37.8 & 38.4 \\
URL Count & 4.0 & 3.9 & 3.5 & 3.8 \\
Has URL (\%) & 99.7 & 100.0 & 100.0 & 99.9 \\
Character Count & 587.6 & 562.6 & 532.5 & 566.9 \\
Sentence Count & 12.9 & 12.6 & 11.8 & 12.6 \\
\midrule
$N$ (posts) & 774 & 660 & 382 & 1,816 \\
\bottomrule
\end{tabular}\\
\footnotesize{Values are means for the highest-rated note per post. Difficulty terciles based on mean survey ratings.}
\end{table}

\clearpage
\begin{table}[H]
\centering
\caption{OLS Regressions: Note Quality on Fact-Check Difficulty}
\label{tab:note_quality_ols}
\begin{tabular}{lcccc}
\toprule
& \multicolumn{2}{c}{Word Count} & \multicolumn{2}{c}{URL Count} \\
\cmidrule(lr){2-3} \cmidrule(lr){4-5}
& (1) & (2) & (3) & (4) \\
\midrule
Fact-Check Difficulty & $-$0.489$^{*}$ & $-$0.490$^{*}$ & $-$0.198$^{**}$ & $-$0.197$^{**}$ \\
& (0.234) & (0.230) & (0.072) & (0.070) \\
log(Likes) & & $-$2.563$^{***}$ & & $-$0.389$^{*}$ \\
& & (0.491) & & (0.155) \\
log(Retweets) & & 2.060$^{***}$ & & 0.150 \\
& & (0.490) & & (0.155) \\
log(Followers) & & 0.918$^{***}$ & & 0.270$^{***}$ \\
& & (0.145) & & (0.041) \\
\midrule
$N$ & 1,816 & 1,816 & 1,816 & 1,816 \\
$R^{2}$ & 0.002 & 0.037 & 0.004 & 0.030 \\
DV Mean & 38.37 & 38.37 & 3.84 & 3.84 \\
\bottomrule
\end{tabular}\\
\footnotesize{Robust standard errors clustered at the post level in parentheses. $^{*}p < .05$; $^{**}p < .01$; $^{***}p < .001$.}
\end{table}

Notes on harder-to-check claims contain fewer URLs ($\beta = -0.20$, $p < .01$) per unit increase in difficulty. Word count is also significantly negative ($\beta = -0.49$, $p < .05$), and this relationship is robust to the inclusion of engagement controls ($\beta = -0.49$, $p < .05$). The effect sizes are small relative to the overall means (38.4 words, 3.8 URLs), and virtually all notes contain at least one URL regardless of difficulty. The dominant pattern documented in the main text, fewer ratings rather than worse ratings, remains the primary explanation for note failure.

\clearpage
\section{Text Features Predicting Perceived Difficulty}\label{app:textfeatures}

What makes a claim subjectively difficult to fact-check? We extracted 15 linguistic features from each tweet's text and used them to predict perceived fact-check difficulty. Table~\ref{tab:text_features} reports the multivariate OLS results with robust standard errors clustered at the post level.

\begin{table}[H]
\centering
\caption{OLS Regression: Text Features Predicting Perceived Fact-Check Difficulty}
\label{tab:text_features}
\begin{tabular}{lcc}
\toprule
Text Feature & $\beta$ & SE \\
\midrule
Word count & 0.005$^{**}$ & 0.002 \\
Has URL & $-$0.021 & 0.045 \\
Number count & 0.025 & 0.015 \\
Has statistic & $-$0.017 & 0.066 \\
Has organization name & $-$0.020 & 0.047 \\
Has named person & $-$0.066 & 0.083 \\
Has hedge language & 0.133 & 0.072 \\
Has definitive language & $-$0.014 & 0.061 \\
Exclamation marks & 0.003 & 0.024 \\
ALL CAPS sequences & $-$0.017 & 0.011 \\
Emoji count & 0.048 & 0.026 \\
Scientific term count & 0.014 & 0.020 \\
Has attribution & $-$0.063 & 0.091 \\
Has quotation marks & 0.034 & 0.051 \\
Question marks & 0.046 & 0.043 \\
\bottomrule
\end{tabular}\\
\footnotesize{DV: mean perceived fact-check difficulty (1--5). Robust standard errors clustered at the post level in parentheses. $N = 3{,}512$. $R^{2} = 0.011$. $^{*}p < .05$; $^{**}p < .01$; $^{***}p < .001$.}
\end{table}

Collectively, observable text features explain only 1.1\% of the variance in perceived difficulty, which is far less than self-rated believability alone (8.2\%). Only word count significantly predicts difficulty ($\beta = 0.005$, $p = .001$), while the presence of hedging language such as ``may,'' ``could,'' or ``reportedly'' trends positive but does not reach significance ($\beta = 0.133$, $p = .065$). The weak explanatory power of surface-level text features suggests that perceived difficulty reflects deeper cognitive assessments of claim verifiability that are not easily captured by automated content analysis.

\clearpage
\section{Cox Proportional Hazards Model}\label{app:cox}

To test whether fact-check difficulty predicts the rate at which posts receive helpful Community Notes over time, we estimated Cox proportional hazards models. All posts enter the risk set at the time of creation. The event is receipt of a first helpful note. Posts that do not receive a helpful note are right-censored at 24 hours.

\begin{table}[H]
\centering
\caption{Cox Proportional Hazards Models: Time to First Helpful Note (24-Hour Censoring)}
\label{tab:cox}
\begin{tabular}{lcc}
\toprule
& Model 1 & Model 2 \\
& (Bivariate) & (With Controls) \\
\midrule
Fact-Check Difficulty & 0.77$^{**}$ & 0.77$^{**}$ \\
& [0.65, 0.92] & [0.65, 0.92] \\
log(Followers) & & 1.31$^{***}$ \\
& & [1.19, 1.44] \\
log(Retweets) & & 1.09 \\
& & [0.84, 1.41] \\
log(Likes) & & 0.92 \\
& & [0.71, 1.19] \\
\midrule
$N$ (posts) & 2,250 & 2,250 \\
Events & 139 & 139 \\
\bottomrule
\end{tabular}\\
\footnotesize{Hazard ratios with 95\% confidence intervals. Event = post currently holds a helpful note. Posts without a helpful note are right-censored at 24 hours. $^{*}p < .05$; $^{**}p < .01$; $^{***}p < .001$.}
\end{table}

Each one-unit increase in perceived fact-check difficulty reduces the hazard of receiving a helpful note by approximately 23\% (HR $= 0.77$, 95\% CI [0.65, 0.92], $p < .01$). Among the subset of posts that do receive helpful notes, difficulty does not predict the speed of note application (OLS: $\beta = -0.08$, $p = .17$), indicating that the barrier operates entirely through the probability of receiving a note rather than delay. Results are robust to alternative censoring windows (Table~\ref{tab:cox_robustness}).

\begin{figure}[H]
\centering
\includegraphics[width=0.85\linewidth]{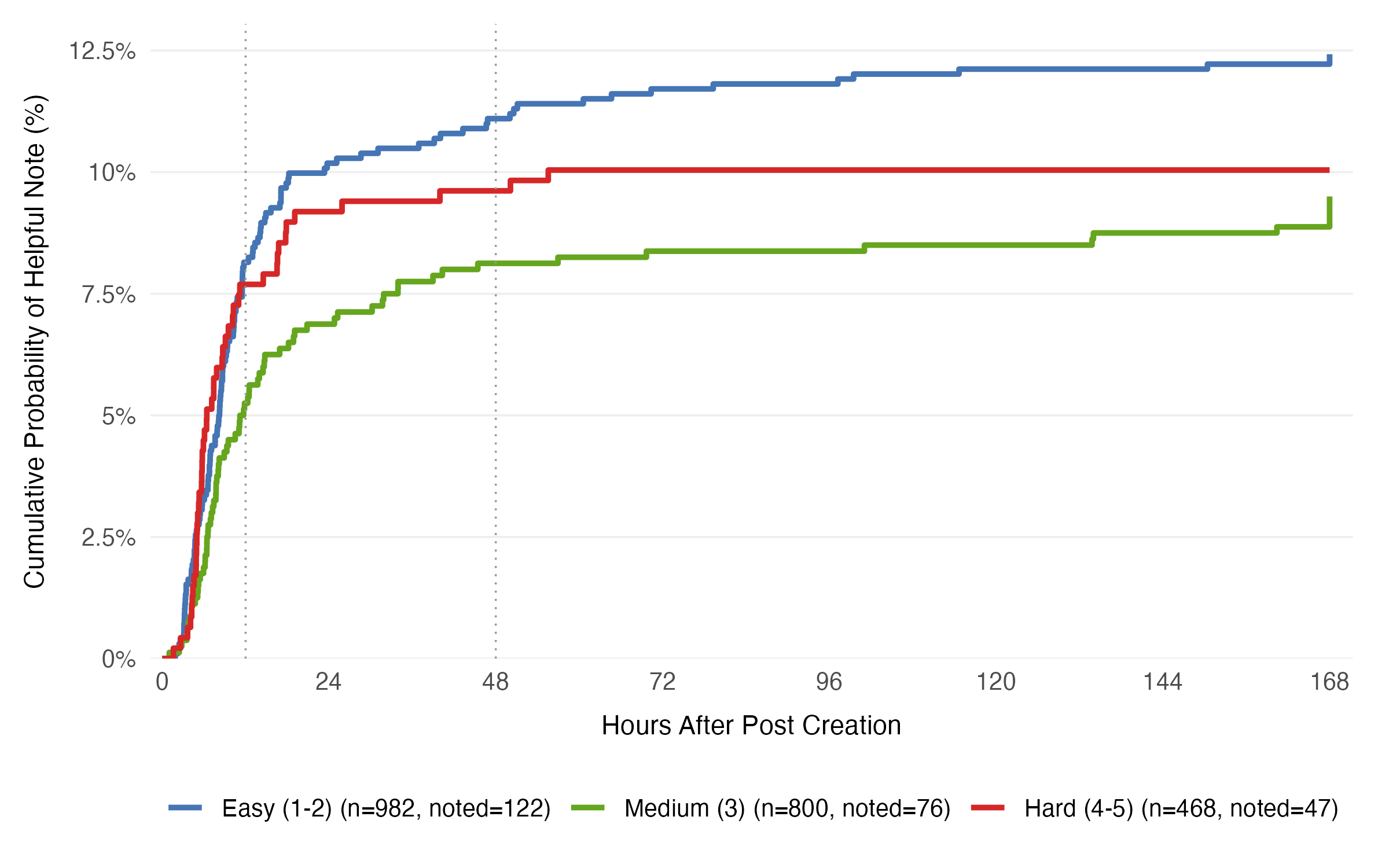}
\caption{Kaplan--Meier Cumulative Incidence of Helpful Community Notes by Difficulty\\
\footnotesize{Note: Step functions show the cumulative probability of receiving at least one helpful Community Note over time, stratified by mean perceived fact-check difficulty. Dotted lines mark 12-hour and 48-hour milestones.}}
\label{fig:survival}
\end{figure}

Figure~\ref{fig:survival} presents Kaplan--Meier cumulative incidence curves stratified by difficulty tercile. Easy-to-check claims (difficulty 1--2) reach higher cumulative note probability than hard claims (difficulty 4--5), with the separation between curves emerging within the first 12--24 hours and persisting thereafter.

\subsection{Cox Robustness: Alternative Censoring Windows}\label{app:cox_robustness}

To assess the sensitivity of the Cox results to the choice of censoring window, Table~\ref{tab:cox_robustness} reports hazard ratios under 24-hour, 48-hour, and 7-day right-censoring. The difficulty penalty is robust across all specifications, with hazard ratios ranging from 0.76 to 0.77.

\begin{table}[H]
\centering
\caption{Cox Proportional Hazards Robustness: Alternative Censoring Windows}
\label{tab:cox_robustness}
\begin{tabular}{lccc}
\toprule
& 24-Hour & 48-Hour & 7-Day \\
\midrule
Fact-Check Difficulty & 0.77$^{**}$ & 0.77$^{**}$ & 0.76$^{***}$ \\
& [0.65, 0.92] & [0.65, 0.91] & [0.64, 0.89] \\
\midrule
$N$ (posts) & 2,250 & 2,250 & 2,250 \\
Events & 139 & 147 & 153 \\
\bottomrule
\end{tabular}\\
\footnotesize{Hazard ratios with 95\% confidence intervals. All models include controls for log(followers), log(retweets), and log(likes). $^{*}p < .05$; $^{**}p < .01$; $^{***}p < .001$.}
\end{table}

\clearpage

\section{Additional Robustness Checks}\label{app:robustness}

\subsection{Post-Level Logistic Regression}

To address potential concerns about the individual-level stacked structure of the main analyses, we aggregated ratings to the post level by averaging across the two raters and re-estimated the primary logistic regression. The post-level results are stronger than the individual-level estimates, consistent with attenuation from measurement error in individual ratings.

\begin{table}[H]
\centering
\caption{Post-Level Logistic Regression: Note Application on Fact-Check Difficulty}
\label{tab:postlevel}
\begin{tabular}{lcc}
\toprule
& (1) Bivariate & (2) With Controls \\
\midrule
Fact-Check Difficulty & 0.75$^{**}$ & 0.75$^{**}$ \\
& [0.63, 0.89] & [0.63, 0.89] \\
log(Followers) & & \cmark \\
log(Retweets) & & \cmark \\
log(Likes) & & \cmark \\
\midrule
$N$ & 2,250 & 2,250 \\
\bottomrule
\end{tabular}\\
\footnotesize{Odds ratios with 95\% confidence intervals. HC1 robust standard errors. $^{*}p < .05$; $^{**}p < .01$; $^{***}p < .001$.}
\end{table}

\subsection{Controlling for Plausibility}

To confirm that fact-check difficulty operates independently of claim plausibility, we re-estimated the primary logistic model including self-rated believability as an additional covariate. Difficulty remains a significant predictor of note failure even after controlling for plausibility, consistent with the mediation analysis in the main text.

\begin{table}[H]
\centering
\caption{Logistic Regression: Note Application Controlling for Plausibility}
\label{tab:plausibility_control}
\begin{tabular}{lcc}
\toprule
& (1) Difficulty Only & (2) + Plausibility \\
\midrule
Fact-Check Difficulty & 0.81$^{***}$ & 0.78$^{**}$ \\
& [0.71, 0.91] & [0.65, 0.93] \\
Self-Rated Believability & & 1.05 \\
& & [0.91, 1.22] \\
\midrule
Controls & \cmark & \cmark \\
$N$ & 3,512 & 3,512 \\
\bottomrule
\end{tabular}\\
\footnotesize{Odds ratios with 95\% confidence intervals. Robust standard errors clustered at post level. $^{*}p < .05$; $^{**}p < .01$; $^{***}p < .001$.}
\end{table}

When both predictors are included, plausibility is not significant ($p = .21$), consistent with H4's mediation finding that plausibility operates through difficulty.

\subsection{Negative Binomial Regressions}

The main text reports OLS regressions of rating counts on difficulty. Because rating counts are non-negative integers with right skew, we re-estimated these models using negative binomial regression. Results, reported as incidence rate ratios (IRRs), confirm the OLS findings.

\begin{table}[H]
\centering
\caption{Negative Binomial Regressions: Rating Counts on Fact-Check Difficulty}\label{tab:negbin}
\begin{tabular}{lccc}
\toprule
& Total Ratings & Helpful Ratings & Not-Helpful Ratings \\
\midrule
Fact-Check Difficulty & 0.93$^{***}$ & 0.91$^{***}$ & 0.96$^{*}$ \\
& [0.90, 0.97] & [0.87, 0.95] & [0.92, 1.00] \\
\midrule
$N$ & 2,505 & 2,505 & 2,505 \\
\bottomrule
\end{tabular}\\
\footnotesize{Incidence rate ratios with 95\% confidence intervals. Controls: log(followers), log(retweets), log(likes). $^{*}p < .05$; $^{**}p < .01$; $^{***}p < .001$.}
\end{table}

Each unit increase in perceived difficulty reduces the expected count of total ratings by 7\% (IRR = 0.93) and helpful ratings by 9\% (IRR = 0.91). The effect on not-helpful ratings is smaller and marginally significant (IRR = 0.96, $p = .025$), consistent with the OLS pattern showing that disengagement is concentrated among helpful-rating contributors.

\subsection{Respondent Fixed Effects}

To rule out the possibility that unobserved respondent characteristics drive our results, we re-estimate the main models with respondent (Prolific ID) fixed effects, identifying effects from within-respondent variation across tweets. Table~\ref{tab:respondent_fe} reports coefficients from linear probability / OLS models estimated via \texttt{fixest::feols()} with standard errors clustered at the tweet level ($N$ = 3,511; 382 respondent FEs).

\begin{table}[H]
\centering
\caption{Main Results With and Without Respondent Fixed Effects}
\label{tab:respondent_fe}
\begin{tabular}{llcccc}
\toprule
& & \multicolumn{2}{c}{Baseline} & \multicolumn{2}{c}{Respondent FE} \\
\cmidrule(lr){3-4} \cmidrule(lr){5-6}
& DV & $\beta$ (SE) & $p$ & $\beta$ (SE) & $p$ \\
\midrule
H1 & Helpful Note Applied & $-$0.013 (0.004) & $<$.001 & $-$0.021 (0.006) & $<$.001 \\
\addlinespace
H2 & Total Ratings & $-$6.49 (2.23) & .004 & $-$7.19 (3.76) & .056 \\
   & Helpful Ratings & $-$5.07 (1.70) & .003 & $-$5.79 (2.94) & .050 \\
\addlinespace
H3 & Self-Rated Believability & 0.322 (0.019) & $<$.001 & 0.109 (0.021) & $<$.001 \\
   & Audience Believability & 0.165 (0.018) & $<$.001 & 0.124 (0.021) & $<$.001 \\
\addlinespace
H4 & Sobel $Z$ (mediation) & \multicolumn{2}{c}{$-$2.86, $p$ = .004} & \multicolumn{2}{c}{$-$2.75, $p$ = .006} \\
\bottomrule
\end{tabular}\\
\footnotesize{All models include controls (log followers, retweets, likes) except H3 (bivariate). SEs clustered at tweet level. $^{*}p < .05$; $^{**}p < .01$; $^{***}p < .001$.}
\end{table}

All hypotheses are supported under respondent FEs. H1 strengthens ($\beta$ moves from $-$0.013 to $-$0.021), indicating the baseline is conservative. H2 point estimates are comparable; the wider standard errors reflect reduced degrees of freedom from absorbing 382 FEs. H3 coefficients attenuate, particularly self-rated believability (0.322 $\to$ 0.109), because the FEs absorb stable respondent-level tendencies in credibility assessment, but the within-respondent relationship remains highly significant. The Sobel test for mediation (H4) remains significant at $p$ = .006.

\clearpage
\section{Robustness of Cross-Tabulation to Binning}\label{app:binning}

\begin{figure}[H]
\centering
\includegraphics[width=\linewidth]{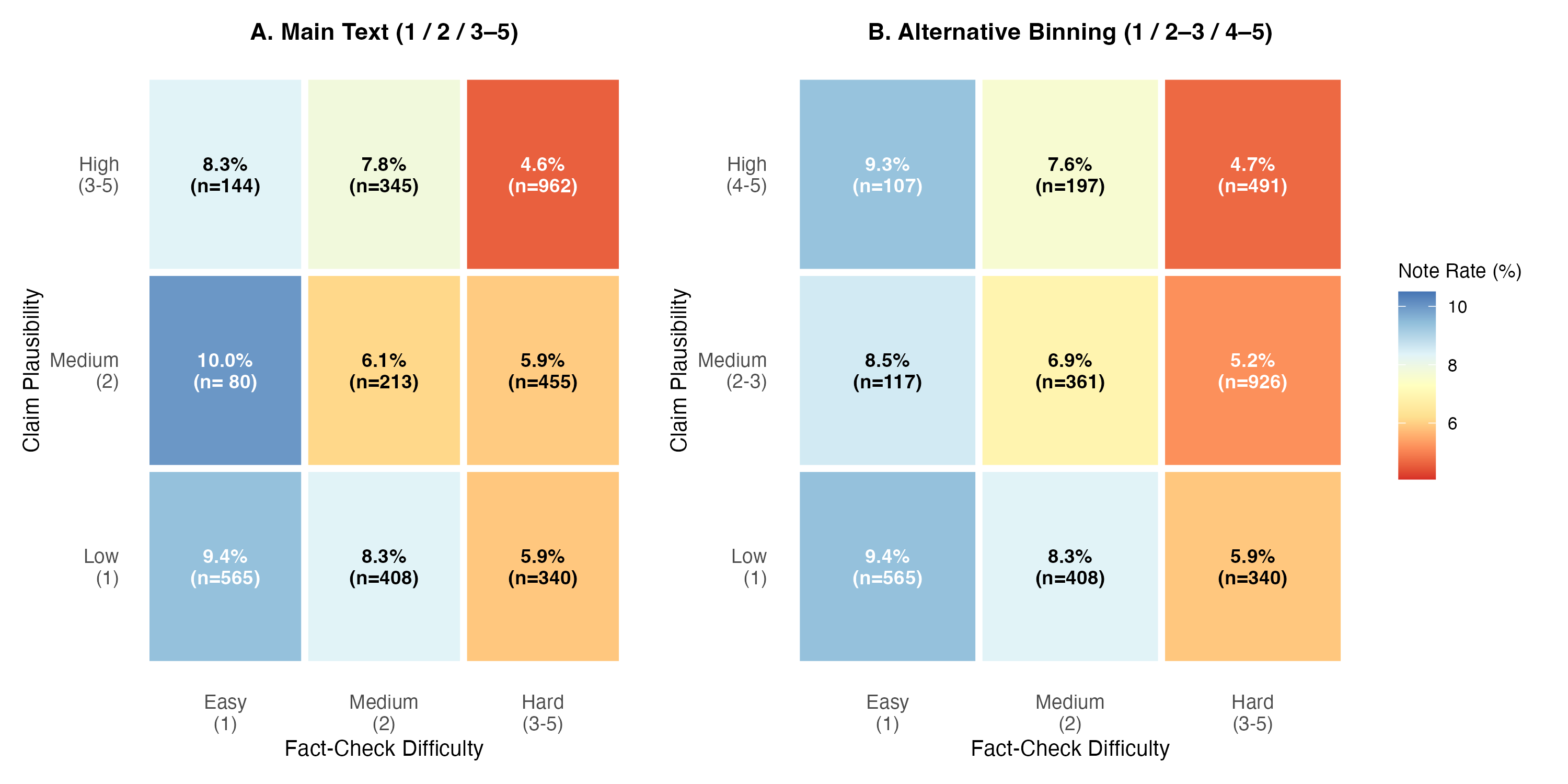}
\caption{Note Application Rate by Plausibility $\times$ Difficulty: Sensitivity to Binning\\
\footnotesize{Note: Color indicates note application rate on a shared scale (blue = more protection, red = less protection). Panel~A: main text bins (1 / 2 / 3--5 for both axes). Panel~B: believability regrouped into approximate terciles (1 / 2--3 / 4--5) while keeping the same difficulty bins.}}
\label{fig:heatmap_binning}
\end{figure}

The cross-tabulation of note application rates by plausibility and difficulty (Figure~4 in the main text) groups individual-level survey responses into three categories per dimension. To ensure the gradient is not an artifact of the specific bin cutpoints, Figure~\ref{fig:heatmap_binning} presents the same analysis under an alternative binning strategy that produces more balanced cell sizes along the believability axis. Panel~A reproduces the main-text bins (1 / 2 / 3--5), while Panel~B regroups believability into approximate terciles (1 / 2--3 / 4--5), yielding splits of 37\%, 40\%, and 23\% of observations rather than 37\%, 21\%, and 41\%. The gradient from high to low note application rates is consistent across both specifications, with the riskiest cell, which contains high plausibility and hard-to-check claims, receiving notes at roughly half the rate of easy, low-plausibility claims.

\clearpage
\section{LLM Fact-Check Validation}\label{app:llm}

To externally validate that perceived fact-check difficulty captures genuine verification complexity rather than an artifact of subjective judgment, we submitted all 2,250 posts in our sample to an automated fact-checking pipeline powered by large language models (LLMs). The pipeline operated in three stages: (1) claim detection (gpt-4.1-mini), which identified whether each tweet contained factual claims and extracted individual assertions; (2) fact-checking (gpt-4.1), which evaluated each claim against web search results and assigned a verdict of ``true,'' ``false,'' or ``uncertain'' with a confidence score; and (3) cascade escalation (gpt-5), which re-evaluated claims that received low-confidence or ``uncertain'' verdicts from the primary model. The pipeline also performed OCR on tweet screenshots containing images. Per-tweet timing and token usage were recorded from the API response metadata.

Of 2,250 posts, 2,169 (96.4\%) were identified as containing at least one factual claim, yielding 4,765 total claims (mean = 2.12 per tweet). GPT judged 72.4\% of posts as ``false,'' 16.9\% as ``true,'' 7.2\% as ``uncertain,'' and 3.6\% as non-factual.

\subsection{Does Perceived Difficulty Predict GPT Fact-Checking Time?}

If perceived difficulty reflects genuine verification complexity, posts rated as harder to check should require more computational effort from the automated system. Table~\ref{tab:llm_timing} reports OLS regressions of log-transformed total processing time on mean perceived difficulty. Posts that survey respondents rated as harder to fact-check took GPT significantly longer to process ($\beta = 0.050$, $p = .018$), and this relationship survives the addition of engagement controls ($\beta = 0.042$, $p = .041$). Descriptively, posts in the hardest difficulty tercile took GPT an average of 77.0 seconds to process compared to 65.9 seconds for the easiest tercile. Harder posts also contained slightly more extractable claims (Easy: 2.12, Medium: 2.23, Hard: 2.30), suggesting that the additional processing time reflects multi-faceted assertions requiring more verification passes.

\begin{table}[H]
\centering
\caption{OLS: GPT Processing Time Regressed on Perceived Fact-Check Difficulty}
\label{tab:llm_timing}
\begin{tabular}{lcc}
\toprule
 & \multicolumn{2}{c}{DV: log(Total Processing Seconds)} \\
\cmidrule(lr){2-3}
 & (1) Bivariate & (2) + Controls \\
\midrule
Fact-Check Difficulty & 0.050$^{*}$ & 0.042$^{*}$ \\
 & (0.021) & (0.021) \\
log(Followers) & & \cmark \\
log(Retweets) & & \cmark \\
log(Likes) & & \cmark \\
\midrule
$N$ & 2,169 & 2,169 \\
$R^2$ & 0.003 & 0.049 \\
\bottomrule
\multicolumn{3}{l}{\footnotesize{$^{*}p<.05$. HC1 robust standard errors in parentheses.}} \\
\end{tabular}
\end{table}

\subsection{Does Perceived Difficulty Predict GPT's Truth Judgment?}

A potential alternative explanation for our main findings is that claims perceived as difficult to check are actually \textit{true}, and that the difficulty reflects the absence of disconfirming evidence rather than genuine verification complexity. If so, the pattern we document might reflect appropriate restraint by Community Notes raters rather than effort aversion. To test this, we examined whether GPT was more likely to judge harder-to-check posts as true. Table~\ref{tab:llm_verdict} reports logistic regressions of a binary indicator for GPT's ``true'' verdict (vs.\ ``false'' or ``uncertain'') on perceived difficulty, as well as OLS regressions of GPT's self-reported confidence.

GPT does not judge harder-to-check posts as more likely to be true. The logistic model yields a clearly null result both without (OR = 1.07, $p = .24$) and with engagement controls (OR = 1.08, $p = .17$). However, GPT is significantly less confident in its verdicts for harder posts ($\beta = -0.007$, $p = .024$), consistent with these claims being genuinely more ambiguous to adjudicate---harder to confirm \textit{or} deny. Importantly, this reduced confidence does not translate into a higher rate of ``true'' verdicts; GPT rates 18.0\% of easy posts as true, 14.9\% of medium posts, and 21.0\% of hard posts, with no monotonic relationship.

\begin{table}[H]
\centering
\caption{GPT Verdict and Confidence Regressed on Perceived Fact-Check Difficulty}
\label{tab:llm_verdict}
\begin{tabular}{lcccc}
\toprule
 & \multicolumn{2}{c}{Panel A: P(True) --- Logistic} & \multicolumn{2}{c}{Panel B: Confidence --- OLS} \\
\cmidrule(lr){2-3} \cmidrule(lr){4-5}
 & (1) & (2) & (3) & (4) \\
\midrule
Fact-Check Difficulty & 0.069 & 0.080 & $-$0.007$^{*}$ & $-$0.006$^{*}$ \\
 & (0.058) & (0.059) & (0.003) & (0.003) \\
{[OR]} & {[1.07]} & {[1.08]} & & \\
Controls & & \cmark & & \cmark \\
\midrule
$N$ & 2,169 & 2,169 & 2,169 & 2,169 \\
\bottomrule
\multicolumn{5}{l}{\footnotesize{$^{*}p<.05$. HC1 robust standard errors in parentheses. Controls: log(followers), log(retweets), log(likes).}} \\
\multicolumn{5}{l}{\footnotesize{``True'' = GPT verdict is ``true''; reference category = ``false'' + ``uncertain.''}} \\
\end{tabular}
\end{table}

\clearpage


\section{Survey Instructions}\label{app:instructions}
The following instructions and questions match the wording and presentation provided to survey participants on Prolific.

\medskip
\noindent\textit{You will evaluate 15 images of posts based on the following questions:}

\medskip
\noindent\textbf{Q1:} Does this post contain a claim about vaccine safety?

\noindent To be about vaccine safety the post must do at least one of the following (not all):
\begin{itemize}
  \item allege that a vaccine itself causes injury, illness, or death;
  \item explicitly tie vaccines to harm or potential harm;
  \item insinuate that vaccines could be harmful
\end{itemize}

\noindent\textbf{Q2:} Is there more than one claim being made?

\noindent\textbf{Q3:} How believable is the provided claim?

\noindent\textbf{Q4:} Based solely on what is provided, how believable do you think others would find the claim?

\noindent\textbf{Q5:} In your opinion, how difficult would it be to fact-check the claim?

\medskip
\noindent\textit{The goal of the study is to assess medical social media claims based on their plausibility/believability. Please answer as honestly as possible.}

\subsection{Response Scales}

\noindent\textbf{Q1.} Does this post contain a claim about vaccine safety?
\begin{itemize}
  \item Yes / No / Unsure / Can't read text
\end{itemize}
\textit{If respondents answered ``No,'' ``Unsure,'' or ``Can't read text,'' they skipped to the next image.}

\noindent\textbf{Q2.} Is there more than one claim being made?
\begin{itemize}
  \item No, just one / Yes, more than one / Unsure
\end{itemize}

\noindent\textbf{Q3.} How believable is the provided claim?
\begin{itemize}
  \item 1 (Not at all believable) to 5 (Very believable)
\end{itemize}

\noindent\textbf{Q4.} How believable do you think others would find the claim?
\begin{itemize}
  \item 1 (Not at all believable) to 5 (Very believable)
\end{itemize}

\noindent\textbf{Q5.} In your opinion, how difficult would it be to fact-check the claim?
\begin{itemize}
  \item 1 (Very easy) to 5 (Very difficult)
\end{itemize}

\end{document}